\documentclass[aps,prr,twocolumn,longbibliography]{revtex4-2}
\usepackage{amsfonts}
\usepackage{amsmath}
\usepackage{graphicx}
\usepackage{dsfont}
\usepackage{diagbox}
\usepackage{array}
\usepackage{amssymb}
\usepackage{bbm}
\usepackage{float}
\usepackage{subfigure}
\usepackage{MnSymbol}
\usepackage{url}
\usepackage[colorlinks=true,
            linkcolor=blue,
            anchorcolor=blue,
            urlcolor=blue,
            citecolor=blue
            ]{hyperref}
\usepackage{xcolor}
\usepackage{booktabs}
\pagecolor{white}
\DeclareMathOperator{\Tr}{Tr}

\usepackage{ulem}
\newcommand\redsout{\bgroup\markoverwith{\textcolor{red}{\rule[0.5ex]{2pt}{0.8pt}}}\ULon}

\begin{document}

\title{Subdiffusive dynamics and critical quantum correlations in a disorder-free localized Kitaev honeycomb model out of equilibrium}
\author{Guo-Yi Zhu}
\email{timeexplorer1991@gmail.com}
\author{Markus Heyl}
\affiliation{Max Planck Institute for the Physics of Complex Systems, N{\"o}thnitzer Stra{\ss}e 38, Dresden 01187, Germany}
\date{\today}

\begin{abstract}
Disorder-free localization has recently emerged as a mechanism for ergodicity breaking in homogeneous lattice gauge theories. In this work we show that this mechanism can lead to unconventional states of quantum matter as the absence of thermalization lifts constraints imposed by equilibrium statistical physics. We study a Kitaev honeycomb model in a skew magnetic field subject to a quantum quench from a fully polarized initial product state and observe nonergodic dynamics as a consequence of disorder-free localization. We find that the system exhibits a subballistic power-law entanglement growth and quantum correlation spreading, which is otherwise typically associated with thermalizing systems. In the asymptotic steady state the Kitaev model develops volume-law entanglement and power-law decaying dimer quantum correlations even at a finite energy density. Our work sheds light onto the potential for disorder-free localized lattice gauge theories to realize quantum states in two dimensions with properties beyond what is possible in an equilibrium context.
\end{abstract}

\maketitle

\section{Introduction}
It is the general expectation that realistic isolated quantum many-body systems driven out of equilibrium eventually thermalize such that the relaxed long-time steady states become locally indistinguishable from thermal ensembles~\cite{Deutsch1991,Srednicki1994,Rigol2007,Eisert2014,Deutsch2018,Mitra2017}. Two types of celebrated exceptions beyond this paradigm are quantum integrable models~\cite{Calabrese2005, Calabrese2006, Alba2016, Moore2017,Calabrese2020} and the Anderson or many-body localization (MBL) mechanism imposed by strong disorder~\cite{Nandkishore2014, Abanin2018}.
In two dimensions, the exploration of ergodicity breaking dynamics in interacting systems remains a challenge especially in view of the argued instability of MBL in two dimensions~\cite{Potirniche2018}. 
Recent years have witnessed a new type of mechanism for nonergodic dynamics unique to lattice gauge theories where static local gauge charge or flux serves as a source for an effective internal disorder~\cite{Smith2017,Brenes2017,Yarloo2017}.
Importantly, this so-called disorder-free localization scenario does not rely on breaking translational invariance and can even occur in interacting two dimensional models~\cite{Smith2018,Karpov2020}, opening up a promising route targeting the challenge of realizing quantum states in two dimensional nonergodic systems with properties beyond any equilibrium counterpart.
In this work we show that the Kitaev honeycomb model driven to highly-excited states by a nonequilibrium quantum quench enters a peculiar disorder-free localized phase exhibiting subdiffusive dynamics towards a critical state exhibiting an algebraically decaying dimer quantum correlation function.
Specifically, we investigate the nonequilibrium dynamics in the Kitaev honeycomb model in a weak skew magnetic field starting from a spin polarized initial state. The problem can be mapped to a weakly interacting Majorana fermion model coupled to a static $\mathbb{Z}_2$ gauge field~\cite{Kitaev2005}, which for the considered dynamics becomes effectively disordered. 
Although a number of previous works have considered the intertwined physics between fermion and flux in the Kitaev model~\cite{Knolle2013,Song2016,Nasu2017,Metavitsiadis2017,Gohlke2018,Knolle2018,Gohlke2017,Rademaker2017,Nasu2019a}, the central open question has remained as to whether this model can break ergodicity and can potentially host non-thermal quantum order.
In the noninteracting limit, we find that the gauge flux disorder localizes most of the Majorana fermions but fails to freeze the metallic and critical modes, leading to the observed subdiffusive dynamics although the system is overall nonergodic~\cite{Altman2014,Vosk2014,Luitz2015,Luitz2017,Lezama2019}. We identify the subdiffusive dynamics in both an algebraic spread of quantum correlations and the power-law growth of entanglement. At late times, the system relaxes to a steady state with dimer quantum correlation functions decaying algebraically in space, which is characteristic of quasi-long range order not accessible in thermal equilibrium. We argue that this quasi-long range order implies a divergent multipartite entanglement as quantified by the quantum Fisher information. We find evidence that our main findings are robust against the leading order perturbative Majorana fermion interactions induced by the skew magnetic field according to our numerical calculations for up to 128 spins on long time scales.
Our results can be extended to any $\mathbb{Z}_2$ lattice gauge theory coupled to chiral Majorana fermions as long as the gauge flux can be considered static and disordered on the considered time scales.

\section{Model}
\begin{figure}[t] 
   \centering
   \includegraphics[width=\columnwidth]{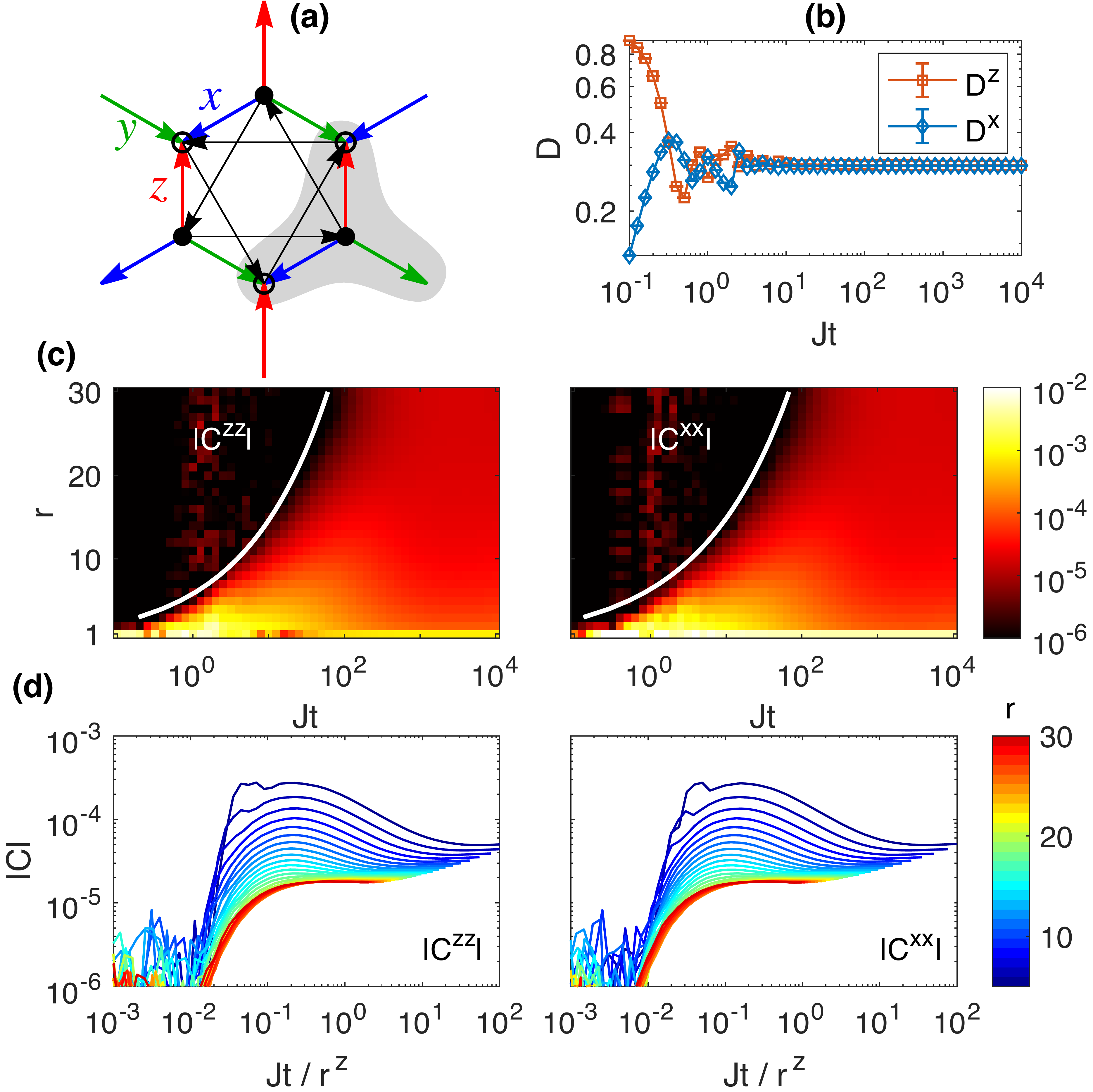} 
   \caption{(a) Model interactions. Arrows indicate Majorana fermion hopping. Four Majorana fermions interact within every Y junction, shaded in gray. (b) $\langle D^{z(x)}(t)\rangle$. (c) $C^{\mu\mu}=\langle D_j^\mu D_{j+r}^\mu\rangle_c$. The propagating wave-front is determined by threshold $|C|\geq 10^{-6}$, subject to a power-law fitting $r\propto (Jt)^{1/z}$ indicated by white lines. $z=2.5(2)$ for $|C^{zz}|$ and $z=2.7(3)$ for $|C^{xx}|$. (d)Collapse of correlation growth at fixed distance with rescaled time. The parameters are $J_z=J_x=J_y=J, \tilde{h}=0.25J$, 10000 disorder samples of system size with $60\times60$ unit cells (7200 spins). }
   \label{fig:light cone}
\end{figure}

The Kitaev model consists of spin-$\frac{1}{2}$ degrees of freedom on the honeycomb lattice with spin-orbital locking Ising interactions $D_j^\mu= -\sigma_j^\mu\sigma_{j+e_\mu}^\mu$, where $j$ labels spin site and $e_{\mu}$ denotes nearest neighbour vector of different orientations $\mu=x,y,z$ (Fig.~\ref{fig:light cone}a). In the presence of a weak $[1 1 1]$ skew magnetic field the Hamiltonian is
\begin{equation} 
\label{eq:HJh}
\hat{H}_K=\sum_j\sum_{\mu=x,y,z}\left(J_\mu \sigma_j^\mu\sigma_{j+e_\mu}^\mu
+ h\sigma_j^\mu\right), \ \ \ h\ll J_\mu.
\end{equation}
For $h=0$ the product $\hat{W}\equiv D^xD^yD^zD^xD^yD^z$ surrounding a hexagon plaquette commutes with $\hat{H}_K$, implying an extensive number of local integrals of motion. 
In the targeted limit $h\ll J_\mu$ we take into account the magnetic field perturbatively to the leading order that preserves these local symmetries~\cite{Kitaev2005}:
\begin{equation}
\hat{H}  =  \sum_{\mu}\sum_{j}J_\mu \sigma_j^\mu\sigma_{j+e_\mu}^\mu
+
\tilde{h} \sum_{(ijk)\in\wedge,\text{Y}} \sigma_i^x\sigma_j^y\sigma_k^z,
\end{equation}
where $\tilde{h}\propto h^3/J^2$. The perturbative interaction acts on three spins that live on any wedges $\wedge$, or the end of any Y junction. Either by introducing the gauge redundancy~\cite{Kitaev2005} and fixing the gauge, or by a Jordan-Wigner transformation~\cite{Feng2006,Lee2007}, one can map $\hat{H}$ onto an interacting Majorana fermion minimally coupled with $\mathbb{Z}_2$ gauge field on the links:
\begin{equation}
\label{eq:HAlpha}
\begin{aligned}
\hat{H} &= \sum_{\langle j \rightarrow l \rangle} J_\mu i u_{j,l} \beta_j \alpha_l
+
\tilde{h} \sum_{\llangle j \rightarrow l \rrangle} i u_{j,k} u_{k,l} \left( \alpha_j \alpha_l + \beta_j \beta_l \right)\\
& + \tilde{h} \sum_{\text{Y}} u_{i,j}u_{i,k}u_{i,l} \left(\beta_i\alpha_j\alpha_k\alpha_l - \alpha_i\beta_j\beta_k\beta_l\right),
\end{aligned}
\end{equation}
where $\alpha(\beta)$ denotes Majorana fermion on $A(B)$ sublattice marked with open(close) circle in Fig.~\ref{fig:light cone}a. The static gauge field on link is pinned to $u_{j,j-e_{x(y)}}=1,\ u_{j,j-e_{z}}=\pm1$. 
$\hat{W}$ on plaquettes are transformed to be locally conserved $\mathbb{Z}_2$ gauge fluxes.
The last four fermion term is a chiral and gauged Majorana Hubbard interaction~\cite{Rahmani2018}, where $j,k,l$ are arranged in a counter-clockwise order around $i$.

\section{Quantum quench protocol}
We prepare a simple initial state as a Néel state such that $\sigma^z|\Psi_0\rangle=\pm |\Psi_0\rangle$ on $A/B$ sublattice respectively, which is to be evolved by $\hat{H}$ later on. In the fermion representation the initial state becomes 
a gauged fermion vacuum coupled to a disordered gauge field background:
\begin{equation}
|\Psi(t)\rangle=\frac{1}{2^{N/2}}\sum_{\{u\}}
e^{-it\hat{H}_{\{u\}}}
|\{u\}\rangle \otimes |\psi_{\{u\}}\rangle,
\end{equation}
where $N$ is the number of unit cells($z$-links), and the Fock state satisfies $iu_{j,j-e_z}\alpha_j\beta_{j-e_z}|\psi_{\{u\}}\rangle=|\psi_{\{u\}}\rangle$. In the sector $\prod_j\sigma_j^z=1$ the anti-periodic boundary condition in spin Hamiltonian is mapped to periodic boundary in Majorana Hamiltonian. 
In this main-text we'll mainly focus on the isotropic coupling $J_x=J_y=J_z\equiv J, \tilde{h}=0.25J$. 

For observables that preserve the gauge field $\hat{O}=\sum_{\{u\}}O_{\{u\}}$~\cite{Smith2017,Brenes2017},
\begin{equation}
\langle \Psi_0 | \hat{O}(t) | \Psi_0 \rangle = \frac{1}{2^N} \sum_{\{u\}} 
\langle \psi_{\{u\}} | e^{it \hat{H}_{\{u\}}}\hat{O}_ {\{u\}}e^{-it \hat{H}_{\{u\}}} | \psi_{\{u\}}\rangle,
\end{equation}
where the average over gauge-field configurations can be performed via Monte-Carlo sampling. The typical $\{u\}$ configuration is random, making the dynamical problem equivalent to Majorana fermions subject to $\mathbb{Z}_2$ gauge ($\pi$) flux disorder, although our model is overall translational invariant~\cite{Smith2017, Brenes2017}. 

Overall we target the description of the nonequilibrium dynamics through a sequence of two steps.
First, we will study in detail the exact solvable point i.e. the noninteracting limit of Eq.~(\ref{eq:HAlpha}), where we find that the system becomes nonergodic due to disorder-free localization, and afterwards explore the influence of interactions.

\section{Exactly solvable point}
When the Majorana interactions are neglected, the model becomes exactly solvable. For each gauge configuration the dynamics is governed by a free Majorana fermion Gaussian Hamiltonian that can be computed efficiently. 
By randomly sampling gauge fields $u=\pm1$ on the $z$-links, we compute the real-time evolution of various physical observables that are natural in both the spin and the fermion language.

First, we consider the spin dimer expectation values $D_j^\mu = iu_{j,j-e_\mu}\alpha_j\beta_{j-e_\mu} (\mu=z,x)$, which relax exponentially fast to the same constant loosing the memory of initial anisotropy (see Fig.~\ref{fig:light cone}b). The observables along $x$ and $y$ directions can be related by mirror symmetry. 
As we will show, the dimer quantum correlation functions
$C^{zz(xx)}(r, t) = \frac{1}{N}\sum_j\langle \Psi(t) | D_j^{z(x)} D_{j+r}^{z(x)} | \Psi(t) \rangle_c$,
exhibit a much slower and intricate dynamics, which quantifies the correlation of gauged local fermion parity. Remarkably, we find that $C^{zz(xx)}(r, t)$ exhibits an algebraic light-cone in spacetime, see Fig.~\ref{fig:light cone}(c), with the the wave-front following a power-law $Jt\propto r^z$ behavior. The dynamical exponent is obtained as $z=2.5(2)$ for $|C^{zz}|$ while $z=2.7(3)$ for $|C^{xx}|$. Notice that the dynamical exponent $z$ here is associated with information transport instead of particle or energy transport, and $z>1$ signals subdiffusion~\cite{Luitz2016}. In Fig.~\ref{fig:light cone}d we further corroborate this by achieving a data collapse upon rescaling the time axis $Jt / r^z$.
While such subballistic behavior in systems with conventional disorder is typically observed on the ergodic side close to the MBL transition lying between diffusive and glassy limit~\cite{Altman2014,Vosk2014,Luitz2015,Luitz2017,Lezama2019}, here we observe such dynamics for a disorder-free localized model, as we will argue in more detail below.

\begin{figure}[t] 
   \centering
   \includegraphics[width=\columnwidth]{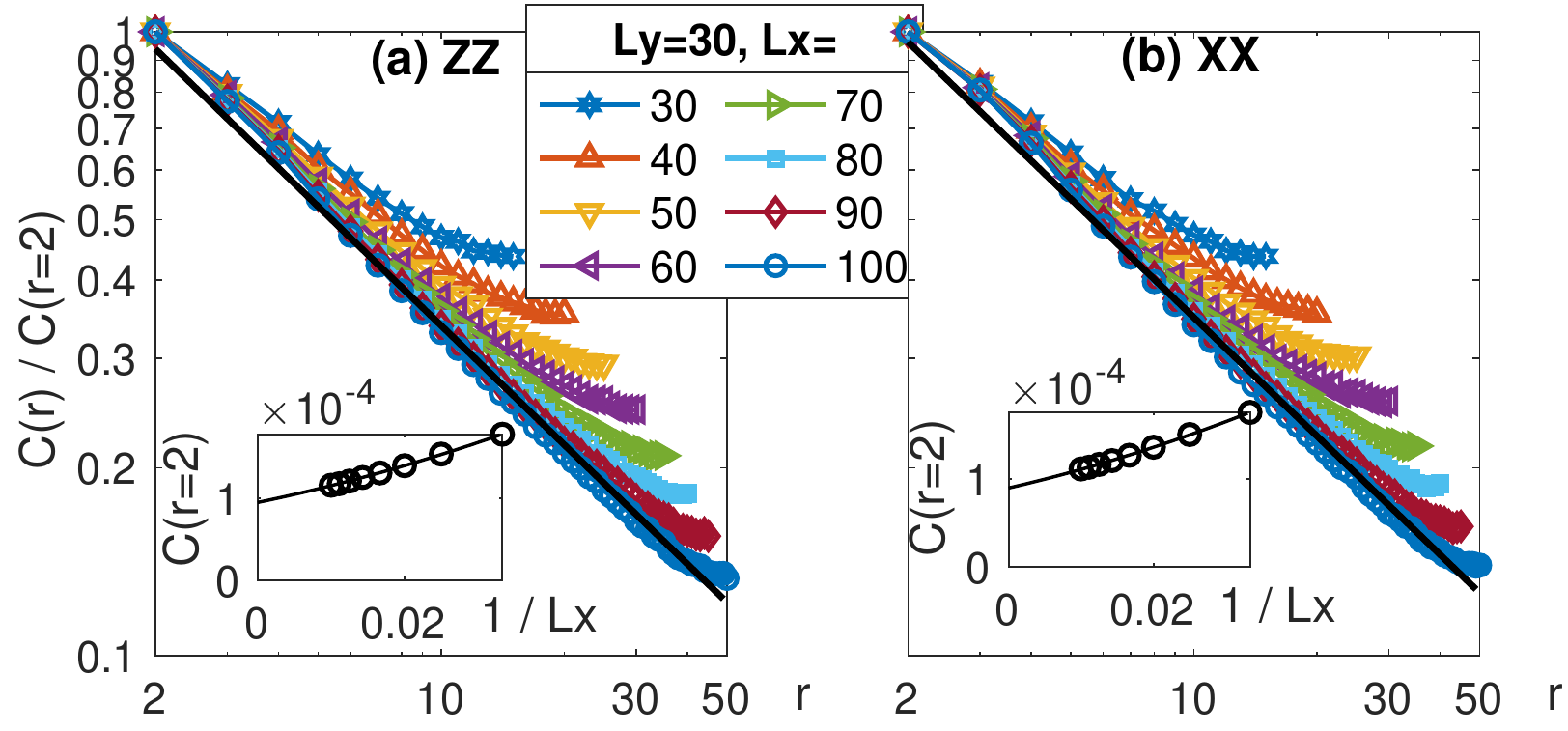} 
   \caption{Dimer correlations in the steady state, averaged over $10^{3.6}\lesssim Jt \lesssim 10^4$. Black lines indicate power-law fitting $r^{-\Delta}$ with $\Delta=0.63(2)$ for both $C^{zz}$ and $C^{xx}$. Insets show correlations at $r=2$ extrapolated to finite value in thermodynamic limit. 
}
   \label{fig:steadyCorr}
\end{figure}

At long times the system settles to a steady state, which, as we find, is of nonergodic critical nature with correlations decaying algebraically in space, as seen in Fig.~\ref{fig:steadyCorr}. We observe that the decay of $C^{zz(xx)}(r, t)$ is consistent with a power-law in space, whose exponent increases for larger system sizes and appears to converge near $0.63(2)$. 
The power-law decaying correlation function in all directions $x,y,z$ are reminiscent of the Kosterlitz-Thouless phase with quasi-long range order~\cite{Kosterlitz1974}, without spontaneously breaking the spin-orbital three-fold rotation symmetry. 
However, even when this symmetry is explicitly broken in the anisotropic regime, we still observe critical correlation~\cite{supplemental}. 
It is the effective disorder that partially inhibits the finite energy density fluctuations and stabilizes the quasi-long range spin dimer order~\cite{Huse2013}.

These critical quantum correlations further have an immediate impact onto the entanglement content of the reached steady state, as the dimer quantum correlation function can be directly linked to a quantum Fisher information density via $f_Q^{zz(xx)}(t) = \sum_r C^{zz(xx)}(r,t)$~\cite{Hyllus2010,Toth2010,Hauke2015}.
Since $C^{zz(xx)}(r,t)\sim r^{-\Delta}$ with $0<\Delta<1$ for $t\to\infty$ we find that $f_Q^{zz(xx)} \sim N^{1-\Delta}$ diverges in the thermodynamic limit.
As a consequence the steady state exhibits strong multipartite entanglement.

\begin{figure}[t] 
   \centering
   \includegraphics[width=\columnwidth]{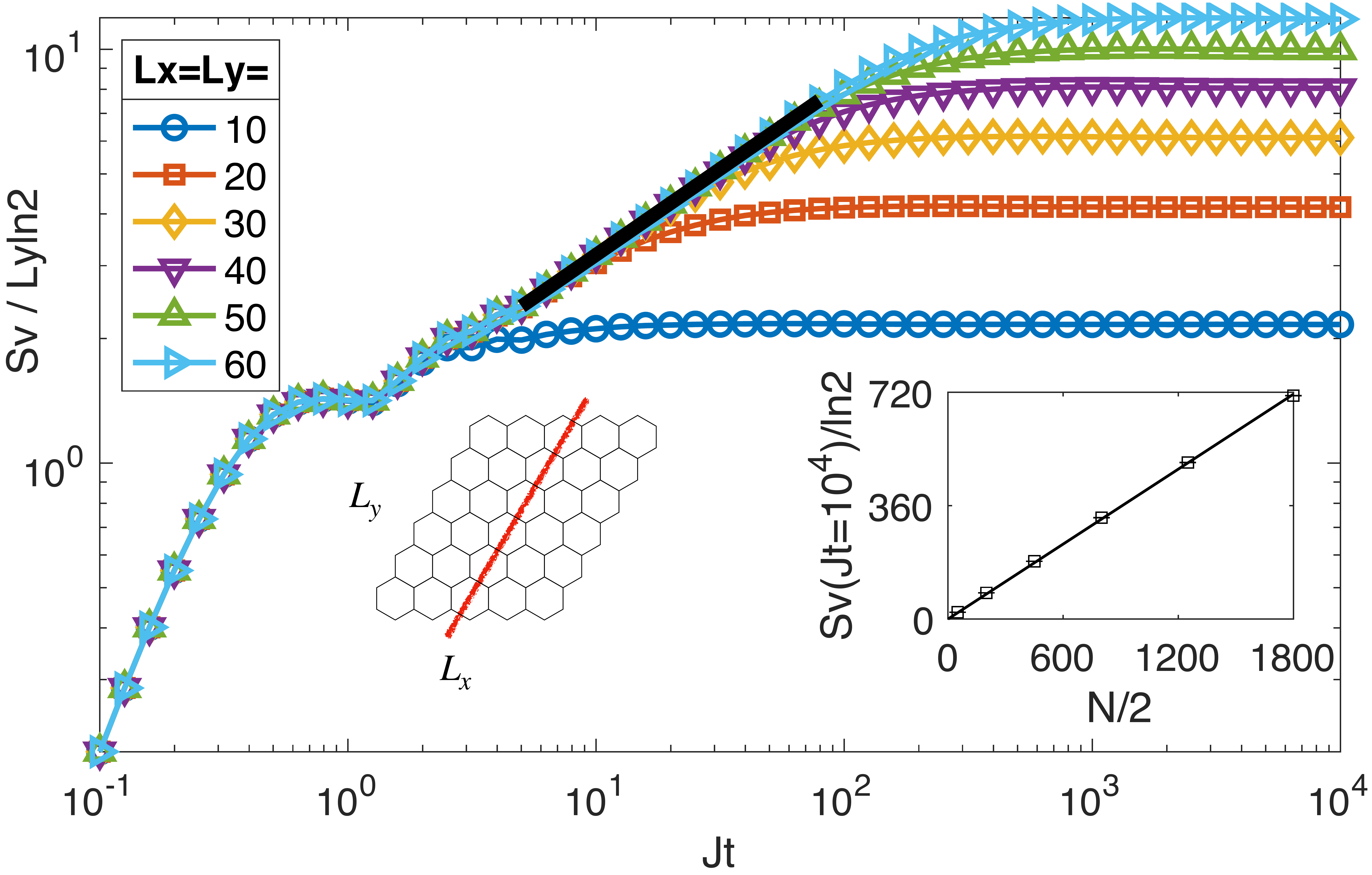} 
   \caption{Projective bipartite entanglement entropy, averaged over 1000 disorder samples. The inset shows the entanglement cut. The black line indicates the power-law fitting $\propto t^{1/z}$ with $z=2.4(1)$. At late time, entropy saturates to a volume law $S_v(Jt=10^4)=0.40(1) N\ln2/2$, as shown in the inset.}
   \label{fig:entanglement}
\end{figure}

For a more detailed quantification of the entanglement properties we further consider the projective bipartite entanglement entropy that serves as an entanglement diagnostics for quantum disentangled liquids~\cite{Grover2013,BenZion2019,Smith2017a}. Namely, we measure the von Neumann entanglement entropy for half of the Majorana fermions, when the gauge field is projected onto the diagonal ensemble:
\begin{equation}
S_v = \frac{1}{2^N}\sum_{\{u\}} S_{\{u\}},\quad
S_{\{u\}} = -  \hat{\rho}_{\{u\}} \ln{\hat{\rho}_{\{u\}}},
\end{equation}
with the reduced density matrix 
\begin{equation}
\hat{\rho}_{\{u\}} = \Tr_{\alpha,\beta \in L} e^{-it\hat{H}_{\{u\}}} |\psi_{\{u\}}\rangle \langle\psi_{\{u\}}| e^{it\hat{H}_{\{u\}}}
\end{equation}
obtained from tracing out Majorana fermions on the left half of the lattice (see inset in Fig.~\ref{fig:entanglement}).
$\hat{\rho}_{\{u\}}$ is a Gaussian operator which can be computed exactly~\cite{Vidal2002,Peschel2002}. 
%
%
Note that while we can compute local observables and correlation functions exactly, the non-projective von Neumann entropy of the Kitaev model is not diagonal with respect to the gauge configurations and therefore cannot be reduced to a free fermion problem, unlike the low order Renyi entropy~\cite{Hart2020}. 
Diagonal entanglement entropies such as the one we consider have been used already for localized systems in other contexts~\cite{Serbyn2013} and give an upper bound on the actual entanglement entropy~\cite{Polkovnikov2008}.
As shown in Fig.~\ref{fig:entanglement}, at early time the entanglement grows with an area law. At a second stage, the entanglement entropy exhibits a further growth according to a subballistic power-law $S\propto t^z$. From a fit to the data we obtain the entanglement dynamical exponent $z=2.4(1)$, which within the accuracy of our simulations aligns with the exponent appearing for the subballistic spreading in $C^{zz(xx)}(r, t)$. In a system of finite size, we find that the entanglement entropy saturates to a volume-law state $S_v\propto L_x L_y$ typical for the highly excited free fermion states~\cite{Lai2014,Lee2014}, as shown in inset of Fig.~\ref{fig:entanglement}.
These numerical findings again highlight the unconventional nonequilibrium dynamics that we observe in the disorder-free localized Kitaev model.

\begin{figure}[h] 
   \centering
   \includegraphics[width=\columnwidth]{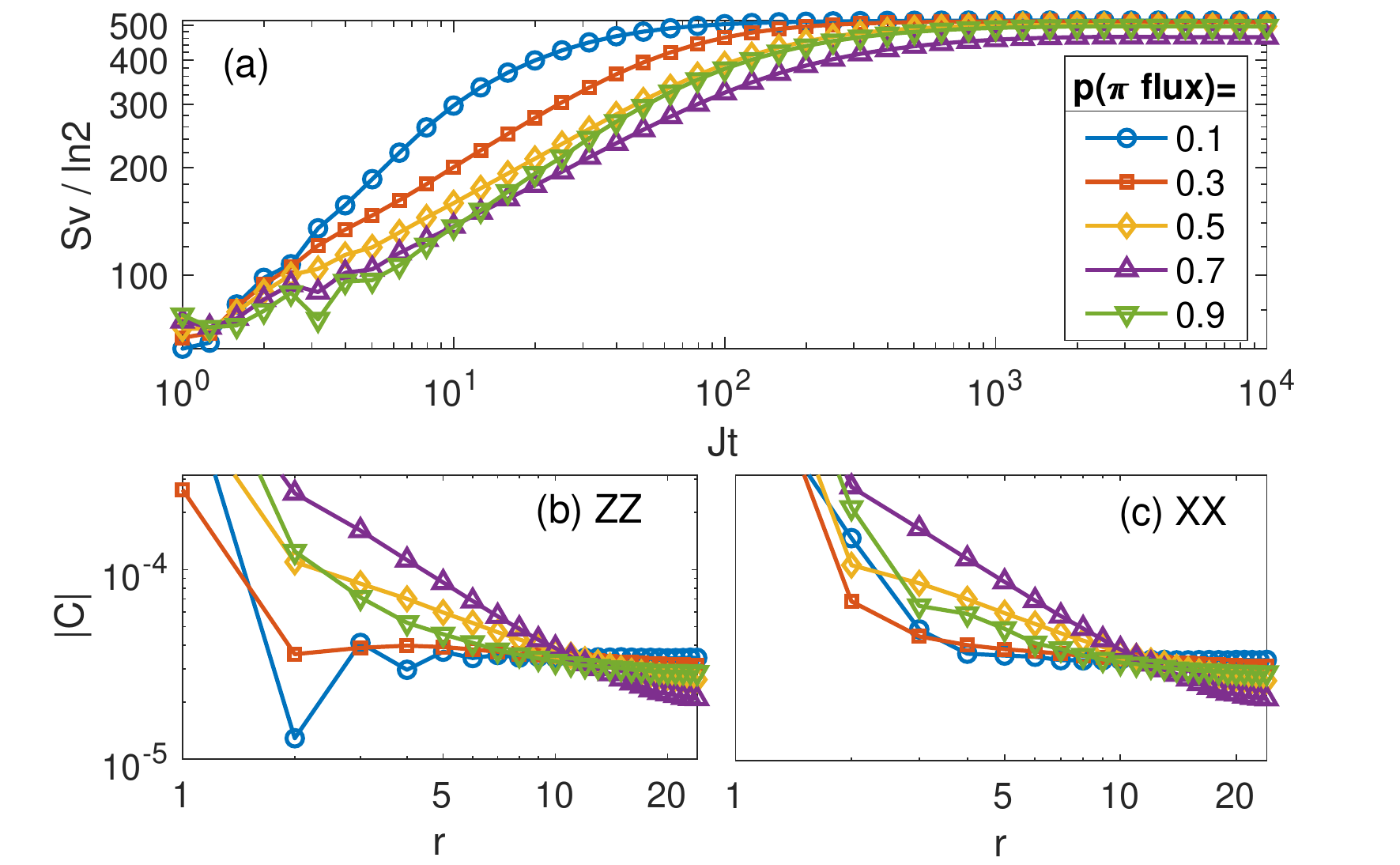} 
   \caption{Tuning the density of $\pi$ fluxes. (a) projective bipartite entanglement entropy of Majorana fermions. (b)(c) Steady state spatial correlation function $|C^{zz(xx)}|$, averaged in the time window $10^3\lesssim Jt \lesssim 10^4$. The parameters are $J_z=J_x=J_y\equiv J, \tilde{h}=0.25J, L_x=L_y=50$, 1000 disorder samples. }
   \label{fig:tuneFlux}
   \end{figure}

One may ask what if we deform the initial state. By tuning the gauge flux density in the initial state by applying an operator $\prod_q(\frac{1}{2}+(\frac{1}{2}-p)\hat{W}_q)$, the exponent $\Delta$ as well as $z$ change continuously, as visible in Fig.~\ref{fig:tuneFlux}, which corroborates the robustness of the critical dynamical phase reminiscent of Kosterlitz-Thouless phase. However, notice that our critical dynamical phase at late-time steady state should be contrasted with the one exhibiting critical initial slip in short-time relaxation~\cite{Janssen1989}. 

\section{Localization analysis}
The peculiar coexistence of subdiffusive dynamics at transient time and the quasi-long-range order at late time implies a subtle localization scenario in behind. Indeed we find a mixture of localized and critical modes from the standard numerical diagnostics~\cite{supplemental} including level spacing statistics~\cite{Oganesyan2006, Devakul2017}, localization length in two dimensions~\cite{MacKinnon,MacKinnon1983, Markos2006}, and Chern number~\cite{Bellissard1994,Kitaev2005,Prodan2010,Bianco2011}. 

\begin{figure}[t] 
   \centering
   \includegraphics[width=\columnwidth]{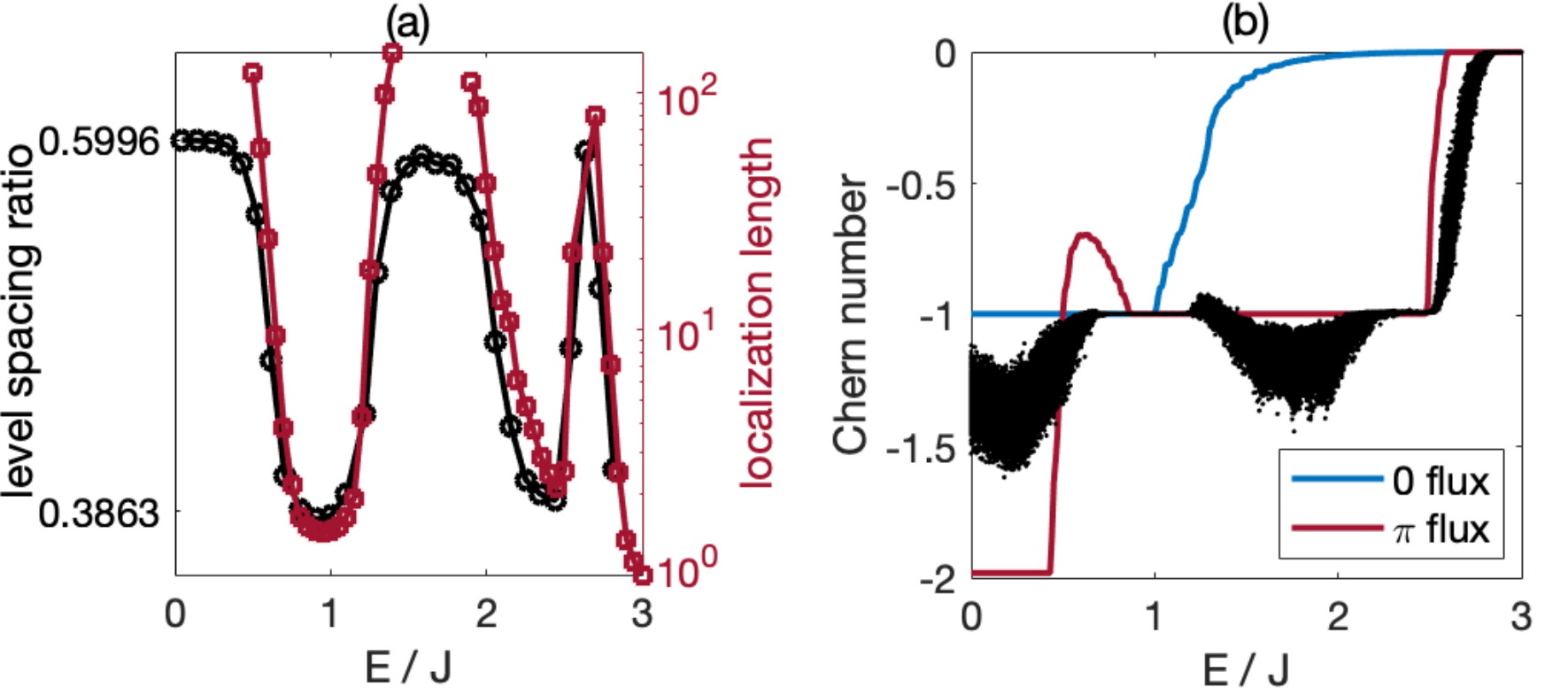} 
   \caption{(a) The left axis shows the level spacing ratio, for system size $L_x=L_y=60$ with 10000 disorder samples. The characteristic value for the Poisson ensemble $2\ln2-1\simeq0.3863$ and the one for Gaussian unitary ensemble $\approx 0.5996$ are indicated. The right axis shows the localization length obtained by one-parameter scaling for a sequence of quasi-1D long stripes for $L_y=8,16,32,64,128$, and $L_x\leq10^6$. (b) Chern number for system size $L_x=L_y=40$. Black dots are for 500 disorder samples, while the blue(red) line is for the zero($\pi$) flux clean system.}
   \label{fig:localization}
\end{figure}

The localization length is calculated by the retarded Green's function using iterative Dyson's equation for a semi-infinite quasi-one-dimensional geometry, followed by a one-parameter-scaling-collapse for varying narrow width. 
As shown in Fig.~\ref{fig:localization}a, the localization length and the level spacing ratio for a finite size system are consistent in showing three energy windows with delocalization tendency. The delocalization at zero energy was known to be responsible for a low-energy Majorana thermal metal state in the class D dirty superconductors~\cite{Read1999,Senthil1999,Chalker2000} or Majorana lattice model~\cite{Laumann2011,Lahtinen2011,Self2018}, which entails logarithmic divergent density of states and weak multi-fractal nature as we numerically verify~\cite{supplemental,Weisse2005}. Intuitively, the low-energy delocalized Majorana mode arises from percolating through an extensive number of resonating Majorana zero modes trapped in $\mathbb{Z}_2$ gauge fluxes in a weak pairing topological superconductor~\cite{Lahtinen2011,Wang2017}. 

To gain more insight into the topology of the fermion, we calculate the Chern number of the fermion eigenstates~\cite{Thouless,Niu,Fukui2005} of varying energy by using the real-space formula based on concept of non-commutative Brillouin zone:
\begin{equation}
C=\frac{2\pi  i}{N}\text{Tr}([P x P, P y P]),
\quad
P(E)=\sum_{\epsilon<-E}|\epsilon\rangle \langle \epsilon|.
\end{equation}
Here $x,y$ are the real-space coordinate operators which label the first quantized orbitals and generate the translation of crystal momenta, $|\epsilon\rangle$ is the single-particle eigenstate of the first quantized Hamiltonian matrix with energy $\epsilon$.
$P(E)$ is the spectral projector where the single-particle mode with energy smaller than $-E$ is occupied, mimicking the Fermi level in the complex fermion system with number conservation. In a fermionic system with only fermion parity conservation, half of the single-particle eigenstates are redundant, so we consider only $E\geq0$. The change of $C(E)$ reveals the Berry flux carried by the fermion mode at the corresponding energy. 
From Fig.~\ref{fig:localization}b, the delocalized mode near $E\simeq 2.5(1) J$ is clearly associated with a topological quantum critical point separating two distinct Chern plateaus, that is robust against perturbation and weak disorder~\cite{Arovas,Evers2007}. As for the energy window $1.5 J \lesssim E \lesssim 2.0 J$, it is unclear whether it would maintain a finite mobility edge or shrink to a singular point or become fully localized in the thermodynamic limit~\cite{Halperin1982,Huo}. 

A final comment is that our result is consistent with the argument that non-Abelian topological phases cannot be fully localized~\cite{Potter2016}.
While disorder tends towards localization, i.e., a divergent dynamical exponent $z\to \infty$, this tendency competes with the metallic and topology induced critical modes favoring ballistic propagation with $z=1$, leading to the observed subdiffusive dynamics with $z>1$.

\section{Beyond exact solvable limit}
\begin{figure}[t] 
   \centering
   \includegraphics[width=\columnwidth]{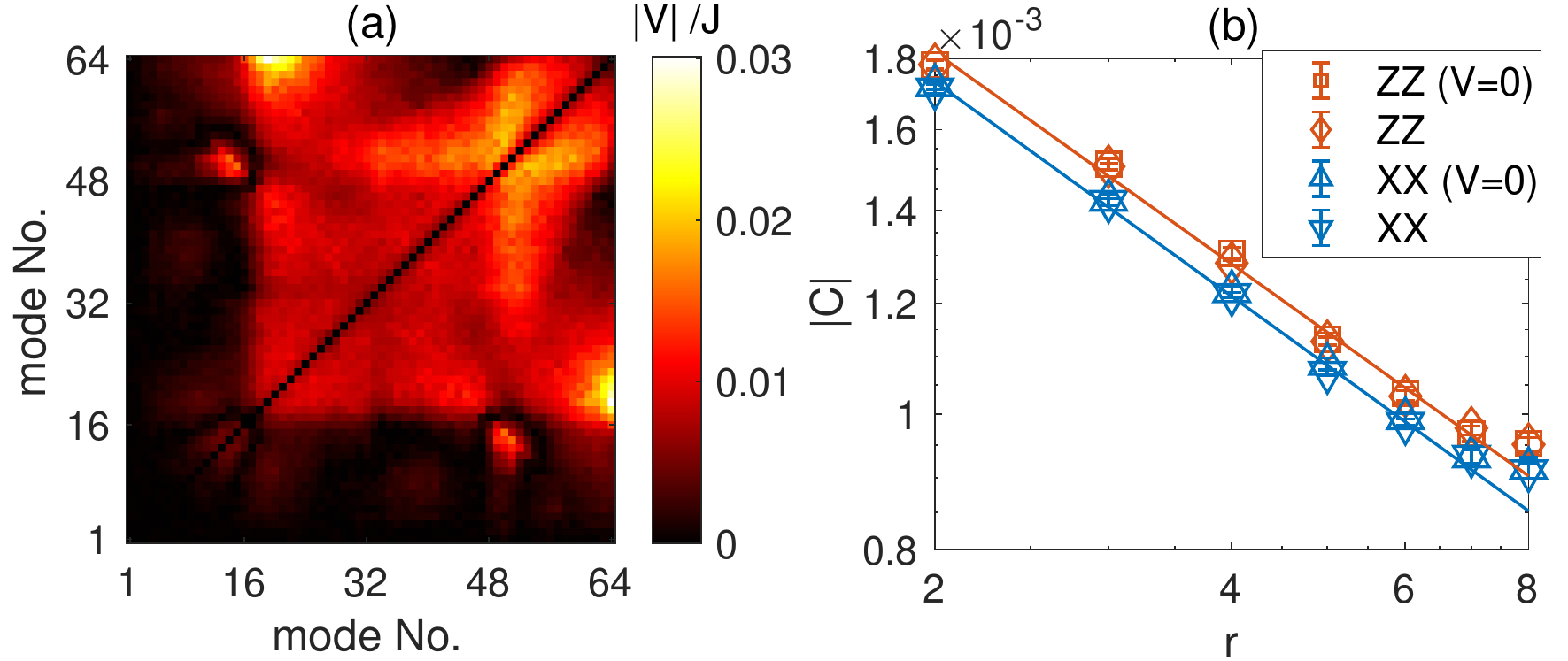}  
   \caption{(a) $V_{mn}$. The mode index corresponds to the energy in ascending order. (b) Correlation functions of steady state, averaged in time window $10^3 \lesssim Jt \lesssim 10^4 $. The power-law-fitting exponent is 0.50(6) for $C^{zz}$, and 0.51(4) for $C^{xx}$. The parameters are $\tilde{h}=0.25J, L_x=16, L_y=4$, 200 disorder samples. The $V=0$ case for comparison takes 10000 disorder samples.}
   \label{fig:interEvol}
\end{figure}

Now we aim to address the robustness of our observations upon the influence of interactions present in Eq.~(\ref{eq:HAlpha}). Here, we will focus on the leading order resonant contributions responsible for an eventual destabilisation, by utilizing the approach introduced in Ref.~\cite{Tomasi2018}, where it has been shown that these resonant contributions can capture the essential non-perturbative effects of interactions such as the logarithmic entanglement growth in MBL phases not only on a qualitative but also on a quantitative level. The Hamiltonian is then expressed in the canonical fermion basis
\begin{equation}
\hat{H}_\gamma=-\sum_{n=1}^{N}\epsilon_n i\gamma'_n\gamma''_n - \frac{1}{4}\sum_{m,n=1}^{N} V_{m,n} \gamma'_m\gamma''_m\gamma'_n\gamma''_n+\cdots,
\end{equation}
where the canonical Majorana fermions $\gamma'$, $\gamma''$ are related to the original local Majorana fermions by an orthogonal transformation obtained from diagonalizing the non-interacting fermion part. 
%
The leading order resonant interaction preserves the parity of the canonical fermion mode $\langle i\gamma_n'\gamma_n'' \rangle$ but induces a dephasing effect, analogous to the $l$-bit theory in MBL systems~\cite{Abanin2018,Nandkishore2014} in which it leads to dramatic non-perturabative effect~\cite{Tomasi2018}. In our case of Kitaev model with weak magnetic field, we also find a special structure for this interaction strength $V_{m,n}$, which endows a hierarchy of dephasing timescales (see Fig.~\ref{fig:interEvol}a). 
It is the key observation in Ref.~\cite{Tomasi2018} that the dynamics of any fermion correlation function can be effectively written as a sum over a number $O(N^4)$ of Gaussian evolution trajectories, schematically abbreviated as
\begin{equation}
\langle \psi (t)| \gamma_m\gamma_n\gamma_p\gamma_q | \psi (t)\rangle =
\sum_{mnpq} C\langle \psi | e^{-it\frac{1}{4} \gamma A_{mnpq} \gamma} \gamma_m\gamma_n\gamma_p\gamma_q | \psi \rangle,
\end{equation}
which can be further factorized into the product of a Loschmidt amplitude quantity and an effective correlation function
~\cite{Robledo2009,Fagotti_2010,Wimmer2011,Klich2002,Klich2014,supplemental}. 
%
As shown in Fig.~\ref{fig:interEvol}d, we calculate the system with $16\times4$ unit cells (128 spins) up to the timescale $Jt\leq 10^4$, where the interacting scenario turns out to collapse with the non-interacting case within numerical accuracy.  
The critical quantum correlations are therefore stable up to this timescale. We estimate the validity of the perturbative approach by statistics of resonances in first-order perturbation theory of the omitted terms. We find that they are off-resonant with probability $\gtrsim99.5\%$ for the considered parameter regime~\cite{supplemental}, above typical thresholds~\cite{Aleiner2009}, so that they can become relevant only via higher-order processes manifesting on longer timescales.

\section{Concluding discussion}
While thermalization may occur eventually for the full Hamiltonian in Eq.~(\ref{eq:HJh}) on long time scales, we emphasize that our findings imply a long intermediate time window with nonergodic behavior leading to exotic quantum dynamics and correlations.
On the other hand, it is tempting to ask whether the exotic dynamics and non-equilibrium quantum order are related to the low-energy non-Abelian Ising topological order~\cite{Kitaev2005,Nayak2007}. However, in the strongly anisotropic coupling regime which in zero temperature exhibit a distinct Abelian $\mathbb{Z}_2$ topological order~\cite{Kitaev1997}, we find similar high energy critical mode and subdiffusive dynamics as well as critical correlation, which goes beyond the low energy universality class~\cite{supplemental}. 
Our findings of subdiffusive dynamics and critical quantum correlations may emerge universally in general $\mathbb{Z}_2$ lattice gauge theories coupled to chiral Majorana matter fermions, provided two essential ingredients: (i)  nontrivial fermion topology; (ii) static disordered gauge flux~\cite{C.Chamon1995,Hatsugai1996,Altland1998,Hart2019}.
Above all, our observation of quasi-long range order with associated divergent multipartite entanglement in a non-equilibrium high-energy steady state marks a concrete first step towards yet unexplored unconventional phase structures in ergodicity breaking two dimensional quantum models. Subdiffusion might be present also in a more general context of Majorana spin liquids as long as the effective $\mathbb{Z}_2$ gauge flux (vison) dynamics is much slower than that of the fermions, which will be a challenging but valuable scope for future research. 
Furthermore, motivated by the recent developments showing that the gauge charge disorder in a 1D unconstrained gauge theories can stabilize a time crystal order~\cite{Russomanno2019}, it would also be interesting to generalize this idea to two dimensions in search of nontrivial spatiotemporal order from a driven Kitaev model~\cite{Po2017,Fulga2019}.
Finally, beyond the conceptual interest, the nonequilibrium quench dynamics can in principle be realized in various quantum architectures including ultracold atoms~\cite{Duan2003,Micheli2006}, superconducting qubits~\cite{You2010,Sameti2019} or topological nanowires~\cite{Kells2014,Sagi2019}, as well as via ultrafast pump-probe techniques in the Kitaev candidate materials at sufficiently low temperatures with suppressed phonon influence~\cite{Zhang2019,Nasu2019a,Knolle19FieldGuide,Motome2020,Ye2020,Metavitsiadis2020}. 

\begin{acknowledgments}
\textit{Acknowledgments}.-- The authors would like to thank G. D. Tomasi, C. Castelnovo and O. Hart for helpful discussions. G.-Y. Zhu also thanks S.-K. Jian and C. Chen for inspiring discussions, and the IT teams in MPIPKS and MPCDF for the technical support in computations.
This project has received funding from the European Research Council (ERC) under the European Union's Horizon 2020 research and innovation programme (grant agreement No. 853443), and M. H. further acknowledges support by the Deutsche Forschungsgemeinschaft (DFG) via the Gottfried Wilhelm Leibniz Prize program.
\end{acknowledgments}

\bibliography{KitaevQuenchDynamics}

\widetext{
\newpage
\appendix




\section{Tuning density of random $\pi$ fluxes}
The quantum quench protocol with a prequench spin product state excites all the allowed gauge configurations like in a thermal ensemble of infinite temperature, where the typical gauge configuration is maximally random. It is also interesting to see the dynamics interpolating between the clean limit and this dirty limit, analogous to tuning the temperature from zero to infinite in a thermal ensemble. In general there could be three related but inequivalent ways of interpolating between the zero gauge flux configuration and the typical maximally random gauge flux configuration. The first way is to endow any specific gauge field configuration with a thermal weight depending on the corresponding Majorana fermion free energy, which is commonly used in Monte Carlo calculations for finite temperature thermal ensemble~\cite{Nasu2017}. The second way is simply to tune the density of $\pi$ link: $p(u=-1)\in [0,0.5]$~\cite{Laumann2011}. But notice that the gauge field on a link is not gauge invariant quantity. The third interpolation way is to tune the average gauge-invariant flux~\cite{Metavitsiadis2017} by applying the spin ring exchange interaction operator $\prod_q (1/2+(1/2-p)\hat{W})$ to the initial state, which is equivalent to tuning the density of $\pi$ flux on a hexagon plaquette from 0 to 0.5. One could even further extend this range to $p\in[0,1]$ to interpolate between the absolute 0 flux and fully packed $\pi$ flux gauge configurations~\cite{Lahtinen2011}, with the dirty limit lying in between. This manner of controlling the density of $\pi$ flux can be realized by deforming the quench protocol and hence the initial state $|\Psi_0(p)\rangle$, where the fermion maintains in a gauged vacuum in each gauge configuration $\langle \Psi_0(p)|iu\alpha\beta|\Psi_0(p)\rangle=1$. 

\begin{figure}[h] 
   \centering
   \includegraphics[width=\columnwidth]{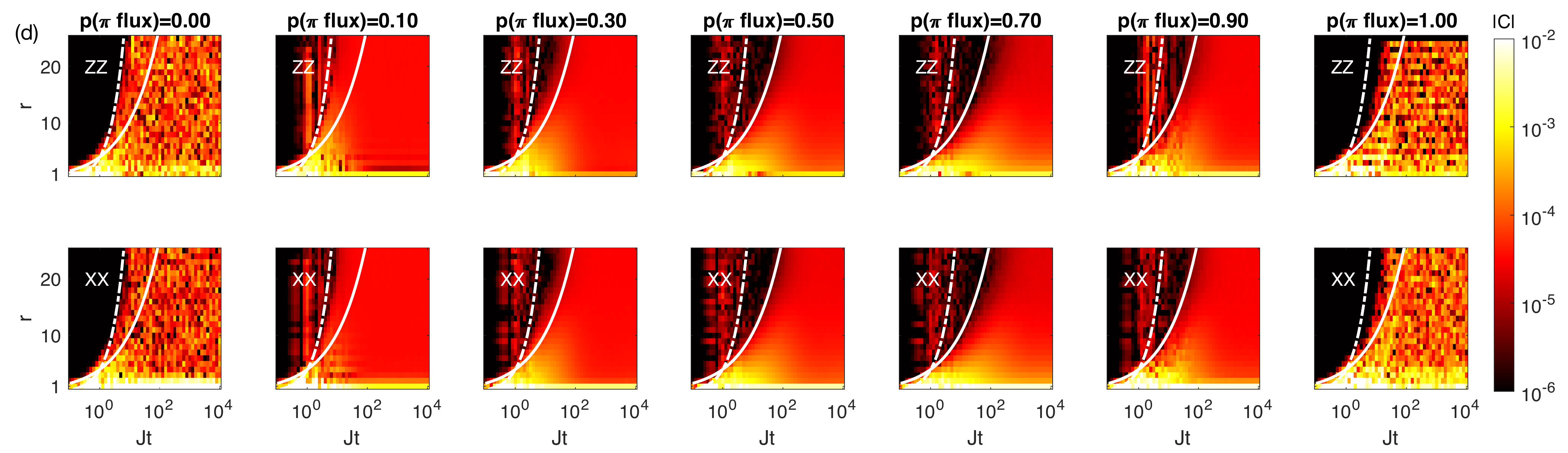} 
   \caption{Correlation light cones by tuning the density of $\pi$ fluxes. Upper(bottom) panel is for $C^{zz(xx)}$. To guide the eyes, white solid line shows subdiffusive propagation $r=4(Jt)^{1/2.4}$ expected for $p=0.5$, while white dot dashed line indicates ballistic propagation $r=4Jt$. Each column stands for a given density of $\pi$ flux. Parameters: $J_z=J_x=J_y\equiv J, \tilde{h}=0.25J, L_x=L_y=50$, 1000 disorder samples. }
   \label{smfig:lightcones}
\end{figure}

As shown in Fig.~\ref{smfig:lightcones}a, at intermediate time, the projected Majorana bipartite entanglement entropy grows algebraically with non-monotonically varying exponent depending on the density of random $\pi$ fluxes. In Fig.~\ref{smfig:lightcones}bc, the late time steady state exhibits algebraic dimer correlation functions in space, with a non-monotonically continuously varying exponent. Overall, the cleaner system with $p$ close to 0 or 1 have faster information propagation, but less prominent critical correlation in steady state. Notice that the maximally random case $p=0.5$ does not yield the slowest entanglement propagation nor the largest correlation exponent, which might be due to the asymmetry between the clean 0 flux sector and $\pi$ flux sector. From this discrete sequence of disorder density, we do not find divergent tendency for the dynamical exponent like in the MBL phase transition, which is consistent with the fact that non-Abelian topological phases of matter cannot be fully localized~\cite{Potter2016}. In Fig.~\ref{smfig:lightcones}d, we show the spatiotemporal profiles of the dimer quantum correlation functions, where a continuously varying light-cone is witnessed. The clean limit without disorder exhibits oscillation inside the lightcone due to the finite size non-interacting nature, which is damped by the onset of disorder. 

\section{Other Majorana fermion correlations}
Throughout the main text we mainly consider the physical observable of spin dimers, which is equivalent to the next nearest neighbouring gauge invariant Majorana fermion bilinear term. This is the simplest and most natural choice in both spin and fermion representation. However, the spin dimer correlation might be much more difficult to measure in experiments than the spin correlation. The single bare spin operator is composed of single matter Majorana fermion and an auxiliary Majorana operator that flips the gauge connection and subsequently the fluxes. Nevertheless, it was shown in Ref.~\cite{Song2016} that the flux-conservation-breaking perturbation could generally dress the spin operator and lead to a contribution of pure matter Majorana fermion bilinear term that preserves the flux. For example, to leading order $\tilde{\sigma}^z \approx \sigma^z + f \sigma^x\sigma^z\sigma^y +\cdots$, where the second term is just the three-spin interaction in an neighbouring wedge, transformed to the next nearest neighbour hopping term of the Majorana fermions on the same sublattice. In low energy, this term contributes to the chiral mass of the Majorana fermion. This pure matter contribution qualitatively changes the low temperature low frequency spin spin correlation function~\cite{Song2016}. Here we also perform a modest calculation for the spreading of (i) correlation between two Majorana fermions on the same sublattice, separated at a distance; (ii) connected correlation between two next-nearest-neighbour-Majorana-fermion-bilinears on the same sublattice. The latter one should contribute to the spin spin correlation that is accessible by experimental probe. As shown in Fig.~\ref{smfig:massCorr}, the spread of this correlation appears to be consistent with the subdiffusive fermion entanglement growth we show in the main-text. In the late time steady state, the correlations also appears to show a power-law signature. Therefore, the subdiffusive dynamics and critical correlation we obtain may leave signatures in the pump-probe experiments for transient spin dynamics. 

\begin{figure}[h] 
   \centering
   \includegraphics[width=0.8\columnwidth]{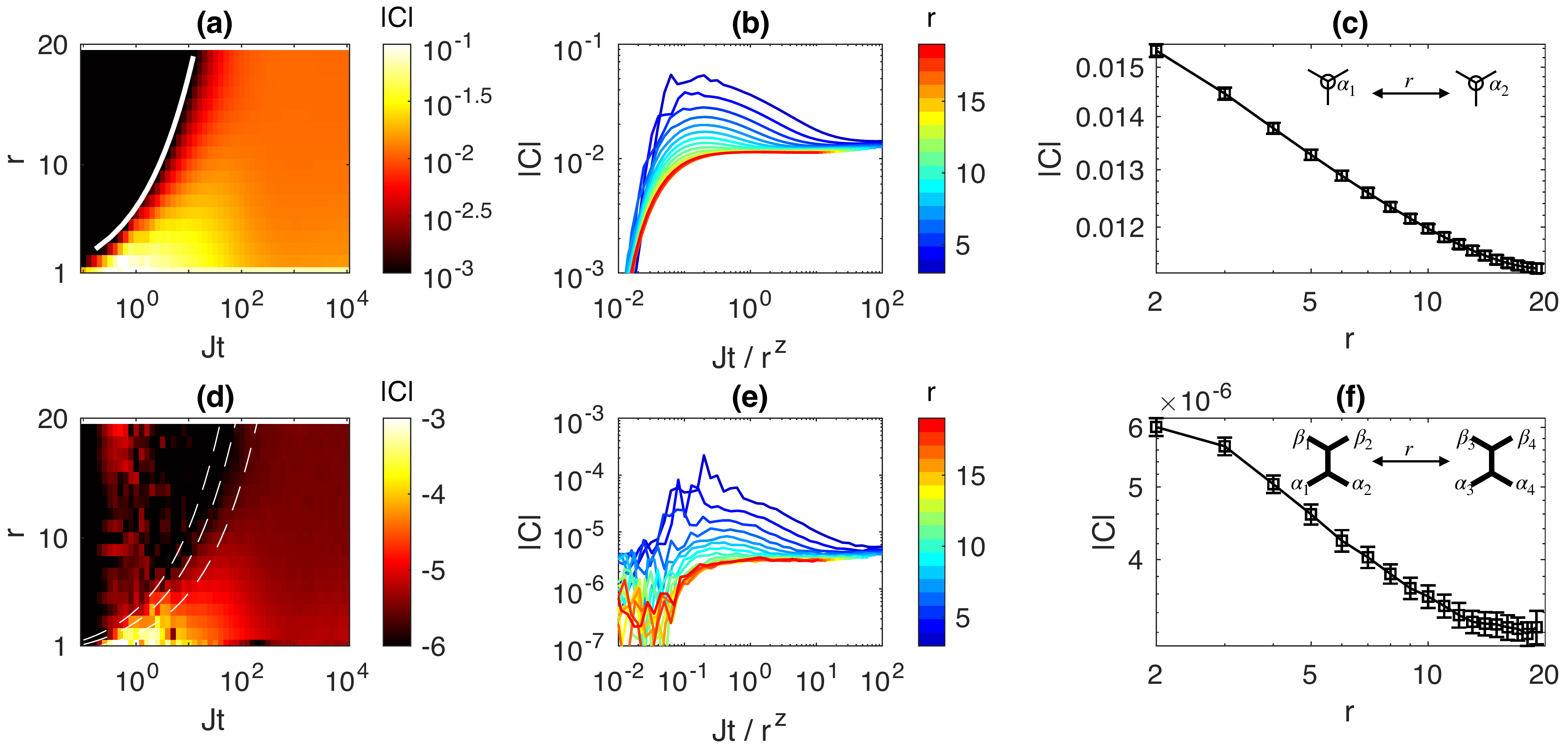} 
   \caption{Spreading of Majorana fermion correlation function (upper panel) and Majorana bilinear correlation function (bottom panel) on the same sublattice. (a) Spatiotemporal tomography of the spreading of Majorana fermion correlation, $|\langle\alpha_j(t)\alpha_{j+r}(t)\rangle|$, being averaged over sites and sublattices and random gauge configurations. By taking a threshold $|C|>10^6$, we fit the propagating front $r\propto t^{1/z}$ where $z=2.3(2)$. (b) Collapse of the growth of correlation function in a rescaled time $Jt/r^z$. (c) Correlation function in steady state, averaged over time window $10^3<Jt<10^4$. Inset shows schematically the fermionic two-point correlation function. (d) Spatiotemporal tomography of the spreading of next-nearest-neighbour-Majorana-bilinear correlation function, $\frac{1}{2^N}\sum_{\{u\}}\langle \psi_u(t)|i\alpha_1\alpha_2i\alpha_3\alpha_4|\psi_u(t)\rangle - \left(\frac{1}{2^N}\sum_{\{u\}}\langle \psi_u(t)|i\alpha_1\alpha_2|\psi_u(t)\rangle\right)^2$, averaged over the sublattices. Despite the large disorder sample size, the statistical fluctuation in small time is still relatively strong compared to the saturated values. Dashed lines show $r=2t^{1/z}$, $r=3t^{1/z}$, $r=4t^{1/z}$ respectively to guide the eyes for the propagating wave-front. (e) Collapse of the growth of correlation function in a rescaled time $Jt/r^z$. (f) Correlation function in steady state, averaged over time window $10^3<Jt<10^4$. Inset shows schematically the  correlation function. Parameters: $J_z=J_x=J_y\equiv J, \tilde{h}=0.25J, L_x=L_y=40$, 10000 disorder samples. }
   \label{smfig:massCorr}
\end{figure}

\section{Anisotropic coupling: beyond low energy universality class}

It is tempting to ask whether the critical dynamical phase we obtain out of a nonequilibrium quantum quench is associated with the low energy topological phases. To answer this question we may tune the Kitaev interaction to be strongly anisotropic, by chooisng $J_z=2J, J_x=J_y=0.5J, \tilde{h}=0.25J$. In this case at zero temperature it belongs to the Abelian $\mathbb{Z}_2$ topological phase, in the same universality class with the celebrated toric code model. In fact, the low energy physics of the anisotropic Kitaev honeycomb model can be mapped to the toric code model~\cite{Kitaev2005}. However, at high energy density the connection between the Kitaev honeycomb model and the toric code model is not \textit{a priori} known. 

The perturbed toric code model was argued to be many-body localized in the presence of strong disorder~\cite{Huse2013}, since it can be dual to the Ising model when charge is absent~\cite{Kitaev1997}. The anisotropic Kitaev honeycomb model in the presence of magnetic field, on the other hand, is found to exhibit subdiffusive spreading of quantum correlation and entanglement, see Fig.~\ref{fig:J2dynamics} and Fig.~\ref{fig:J2entanglement}. Notice that the spreading of dimer correlation is anisotropic, and that along $z$ direction has the same exponent as the spreading of entanglement. Behind that we also find the topological critical fermion modes at high energy for the random flux configurations or $\pi$-flux configuration, even though the zero energy ground state is topological trivial, see Fig.~\ref{fig:J2localization}. It means that unlike the 0 flux sector, the random flux sectors cannot be adiabatically connected to the toric code limit without closing the mobility gap at certain energy. The late-time steady state also exhibits critical dimer correlation along $z$-direction, see Fig.~\ref{fig:J2corr}. These evidences show that the critical dynamics phase goes beyond the low energy effective theory. The topological critical Majorana modes at high energy are responsible for the anomalous subdiffusive dynamics and critical correlation.

\begin{figure}[t] 
   \centering
   \includegraphics[width=.5\columnwidth]{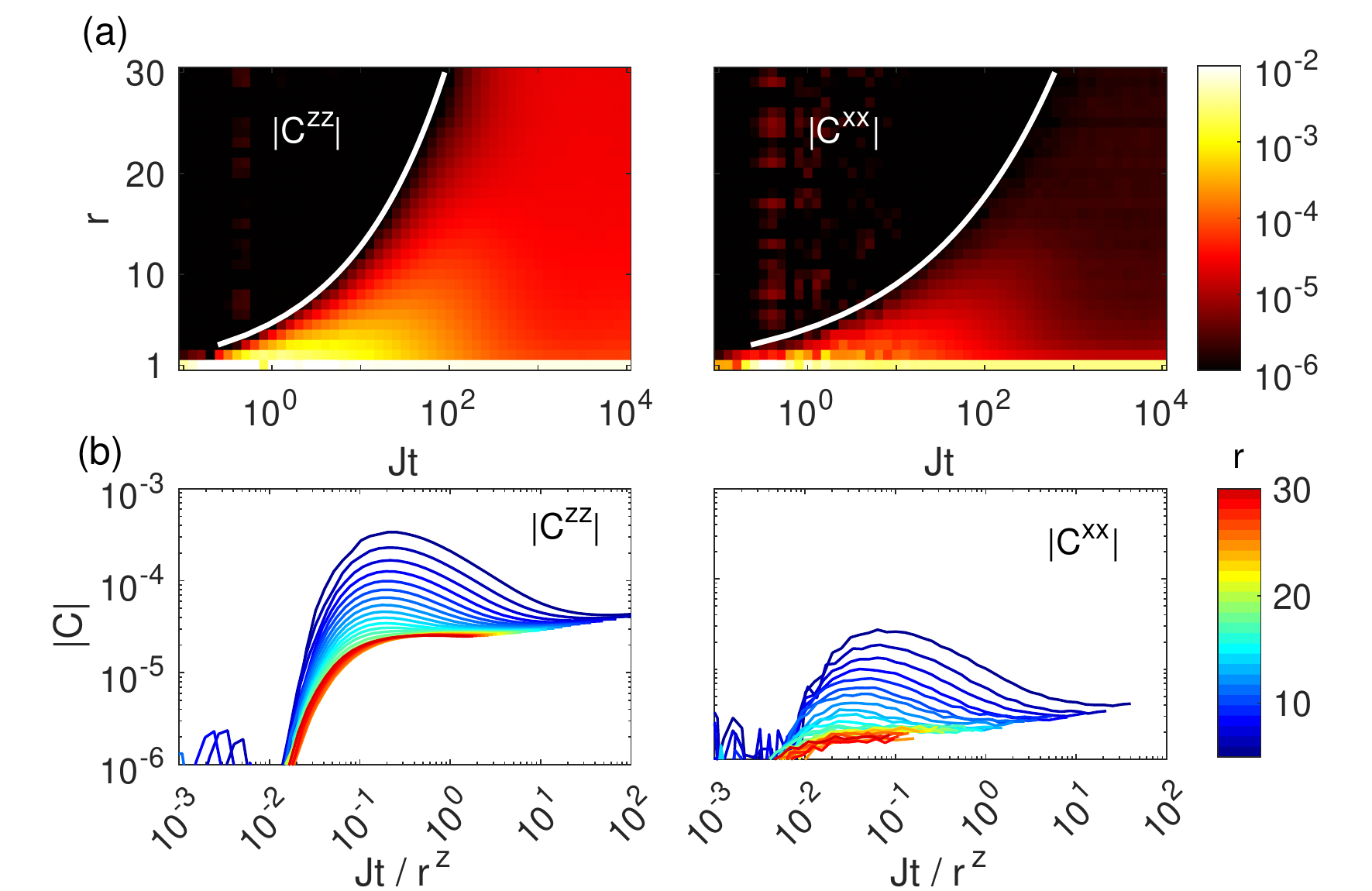} 
   \caption{Anisotropic coupling $J_z=2J, J_x=J_y=0.5J, \tilde{h}=0.25J$. (a) Anisotropic algebraic lightcone of dimer correlation spreading. $L_x=L_y=60$, 1000 disorder samples. White line shows the fitted wave-front using the same threshold as in main text which follows $r\propto (Jt)^{1/z'}$, where $z'=2.6(2)$ for $C^{zz}$ and $z'=3.4(2)$ for $C^{xx}$. (b) Dimer correlation at fixed distances versus rescaled time using $z'$ fitted from entanglement entropy growth below.}
   \label{fig:J2dynamics}
\end{figure}

\begin{figure}[t] 
   \centering
   \includegraphics[width=.5\columnwidth]{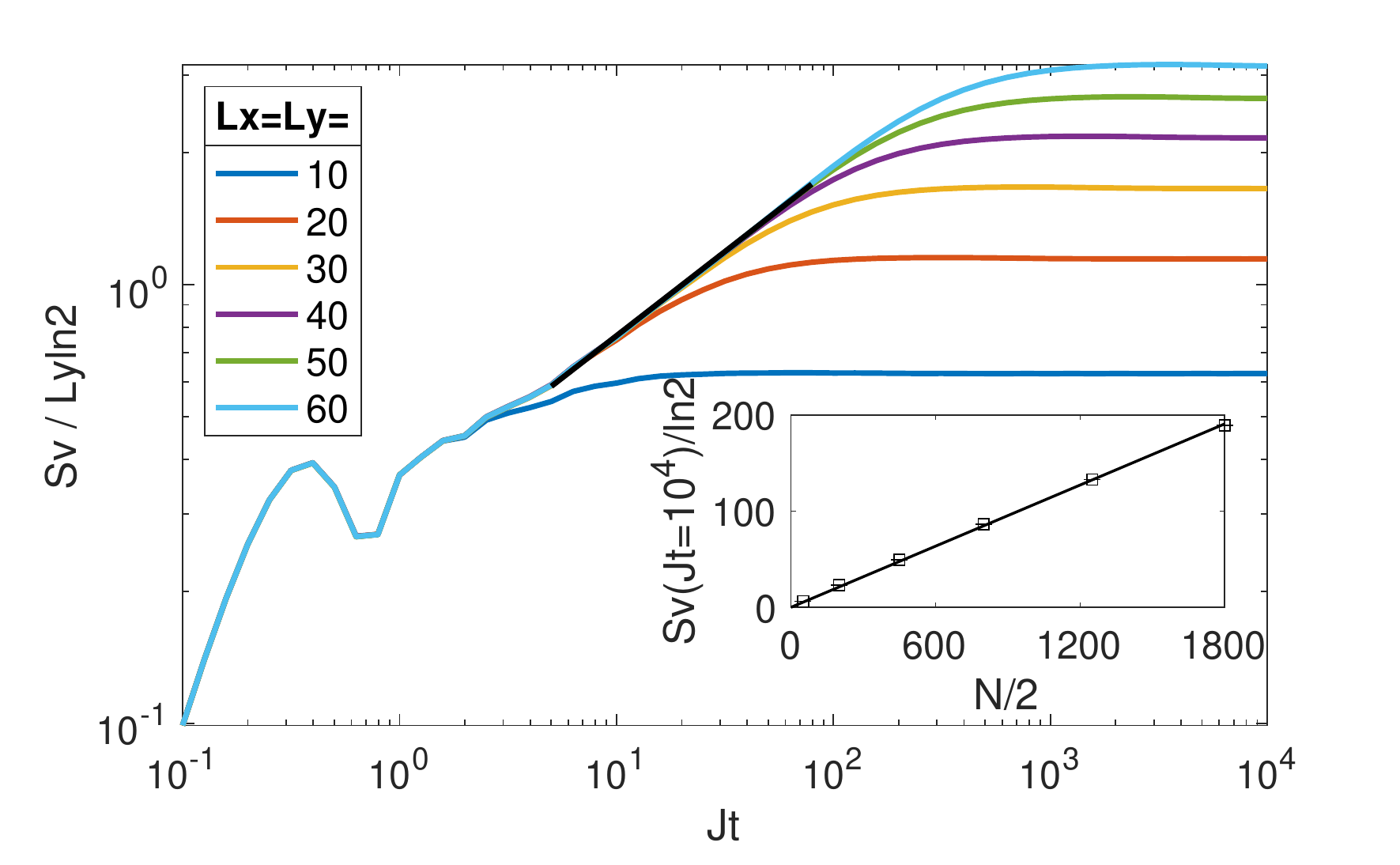} 
   \caption{Anisotropic coupling $J_z=2J, J_x=J_y=0.5J, \tilde{h}=0.25J$. Projective entanglement entropy. Black line is for power-law fit $\propto (Jt)^{1/z'}$ where $z'=2.6(1)$, consistent with the spreading of $C^{zz}$. }
   \label{fig:J2entanglement}
\end{figure}

\begin{figure}[t] 
   \centering
   \includegraphics[width=.5\columnwidth]{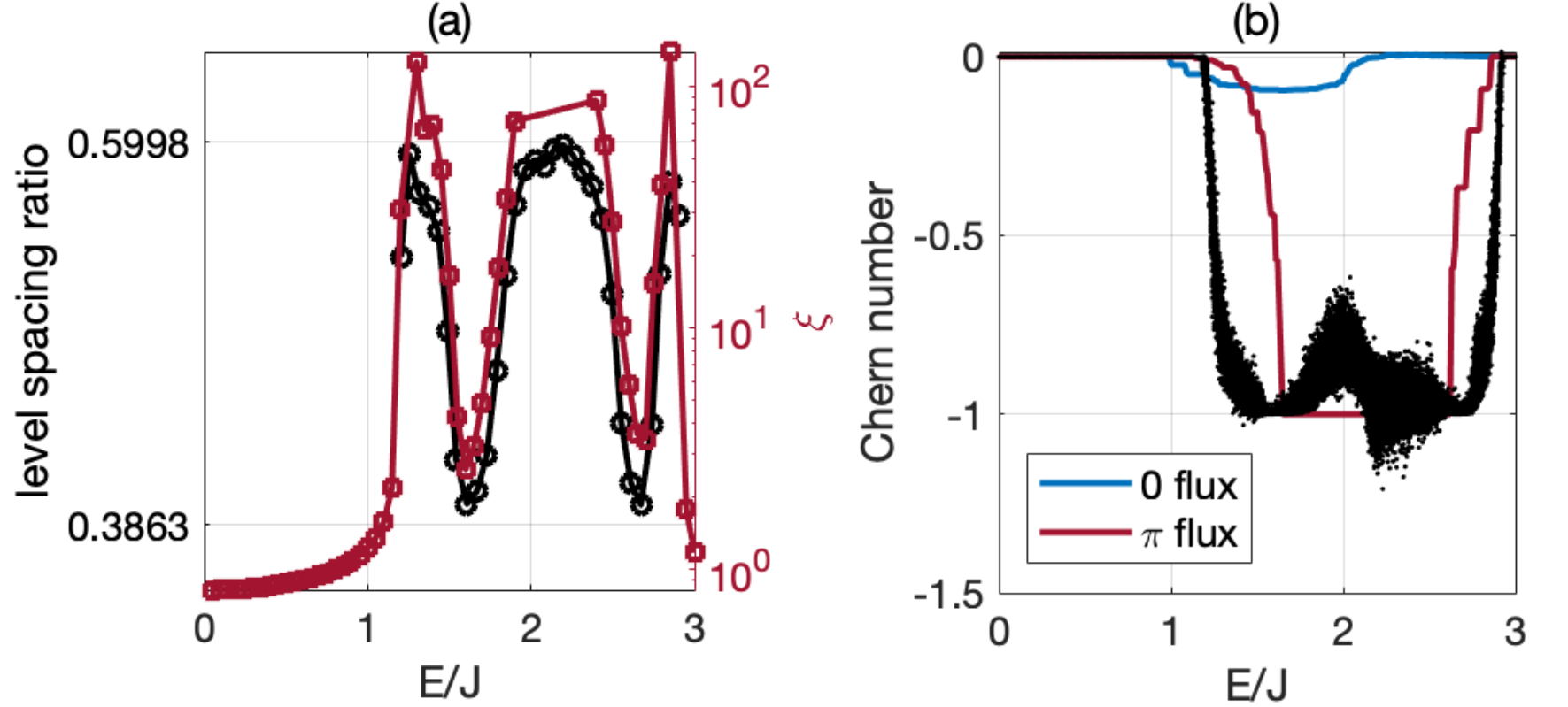} 
   \caption{Anisotropic coupling $J_z=2J, J_x=J_y=0.5J, \tilde{h}=0.25J$. Localization analysis by (a) level spacing ratio and localization length, and (b) Chern number.}
   \label{fig:J2localization}
\end{figure}

\begin{figure}[t] 
   \centering
   \includegraphics[width=.5\columnwidth]{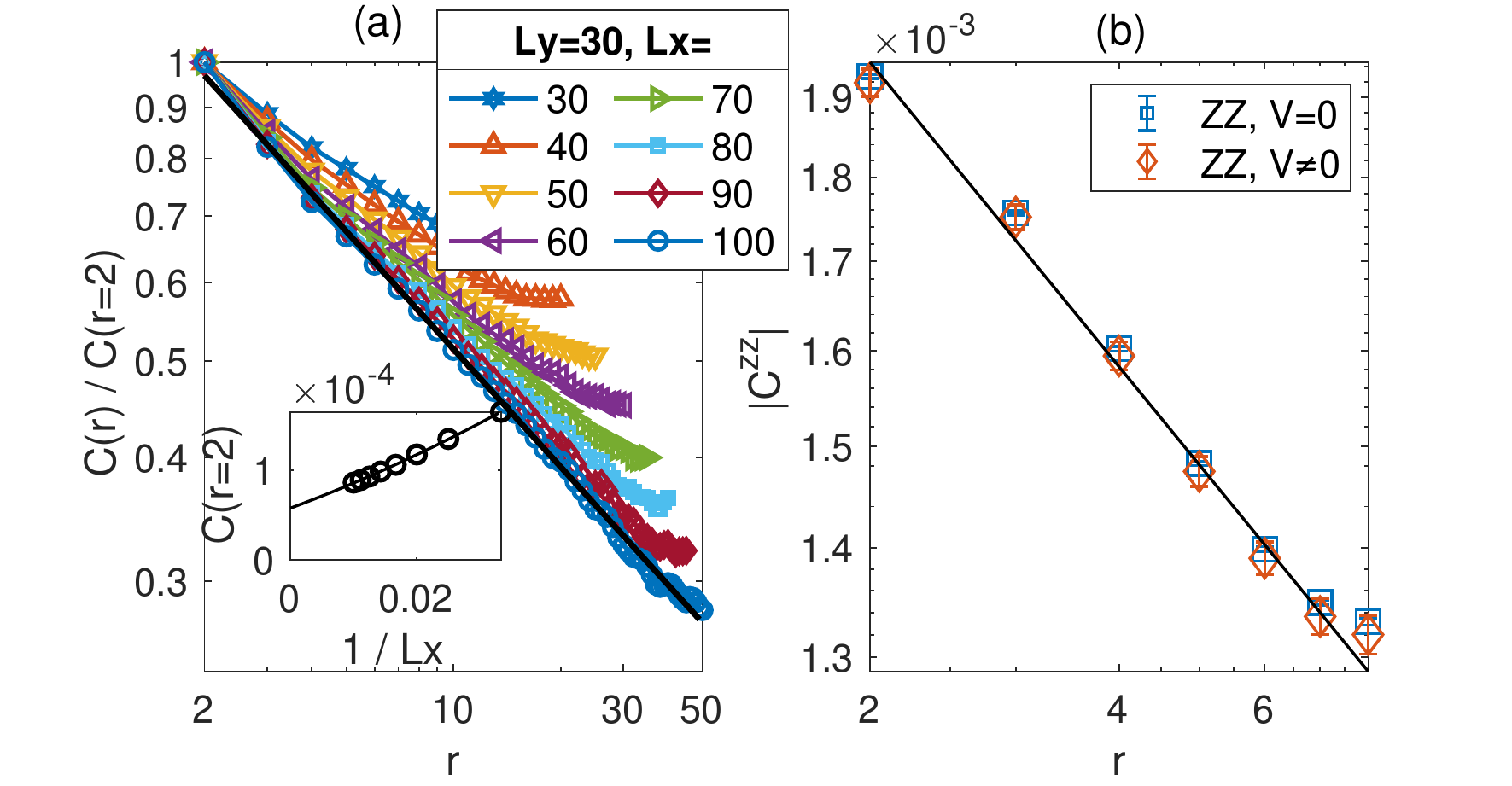} 
   \caption{Anisotropic coupling $J_z=2J, J_x=J_y=0.5J, \tilde{h}=0.25J$. (a) Non-interacting steady state correlation $C^{zz}$ for large system sizes. The black line indicates the fitted power-law behaviour $\propto r^{-\Delta'}$ where $\Delta'=0.39(1)$. (b) Interacting steady state correlation $C^{zz}$, with the non-interacting case in comparison, for system size $16\times4$. Fitted power-law $\propto r^{-\Delta''}$ where $\Delta''=0.30(4)$. }
   \label{fig:J2corr}
\end{figure}

\section{Dynamics in the absence of magnetic field}
For completeness, we also compute the dynamics for the time reversal symmetric Kitaev model in the absence of magnetic field $h=0$, as shown in Fig.~\ref{fig:massless}. In such case, the problem is equivalent to a free Majorana fermion with only nearest neighbour hopping on the honeycomb lattice, that is exposed to a random $\pi$ flux penetrating the plaquette.
\begin{figure}[h] 
   \centering
   \includegraphics[width=0.5\columnwidth]{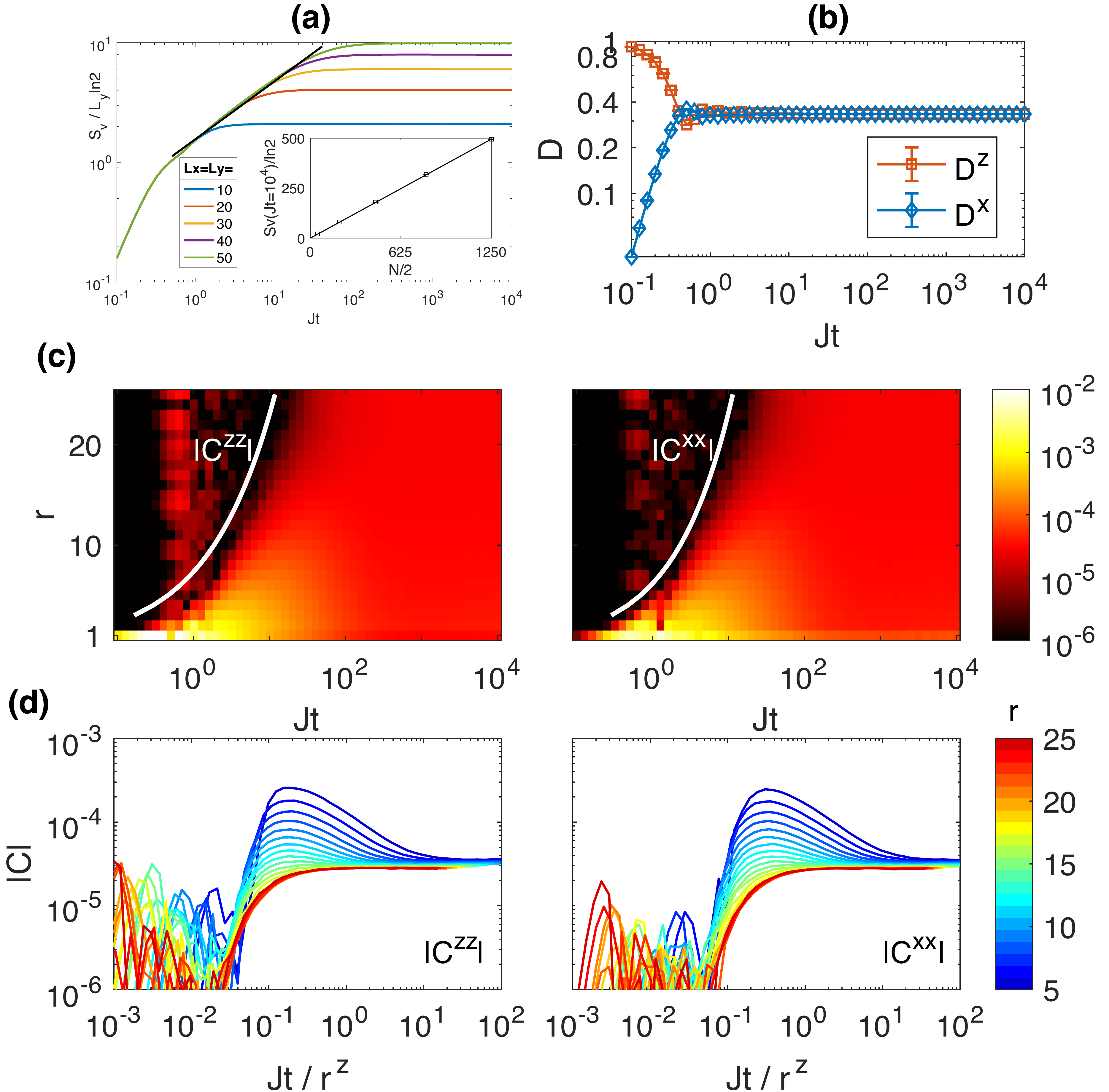} 
   \caption{Numerical results for the gapless Kitaev model without magnetic field $J_z=J_x=J_y=J, h=0$. (a) Projective bipartite entanglement entropy. Black line is for power-law fit $\propto (Jt)^{1/z}$ where $z=2.1(2)$. (b) Dimer expectations. (c) Algebraic lightcone of dimer correlation spreading. $L_x=L_y=50$, 1000 disorder samples. White line shows $r\propto (Jt)^{1/z}$ where $z=2.0(5)$ for $C^{zz}$ and $z=1.7(3)$ for $C^{xx}$. (d) Dimer correlation at fixed distances versus rescaled time using $z$ fitted from the corresponding correlation function.}
   \label{fig:massless}
\end{figure}
Notice that without magnetic field there is not only time reversal symmetry but also the sublattice (chiral) symmetry (sign change to one sublattice changes the sign of the Hamiltonian), which connects the positive energy to negative energy but acts as a unitary symmetry distinct from the anti-unitary particle hole symmetry. The zero energy single particle mode has exact chiral symmetry, and therefore often behaves qualitatively distinct from the other energy states in the presence of disorder. A closely related problem is the gapless Dirac fermion in the presence of random $U(1)$ magnetic flux disorder, see Ref.~\cite{C.Chamon1995,Hatsugai1996,Altland1998}. Here we are dealing with a Majorana version of the Dirac fermion in the presence of $\pi$ flux disorder, with additional particle hole symmetry. As shown in Fig.~\ref{fig:masslessLocalization}, our numerical result suggests a diverging localization length towards zero energy. Besides, there seems to be multiple singular points. 

\begin{figure}[h] 
   \centering
   \includegraphics[width=0.4\columnwidth]{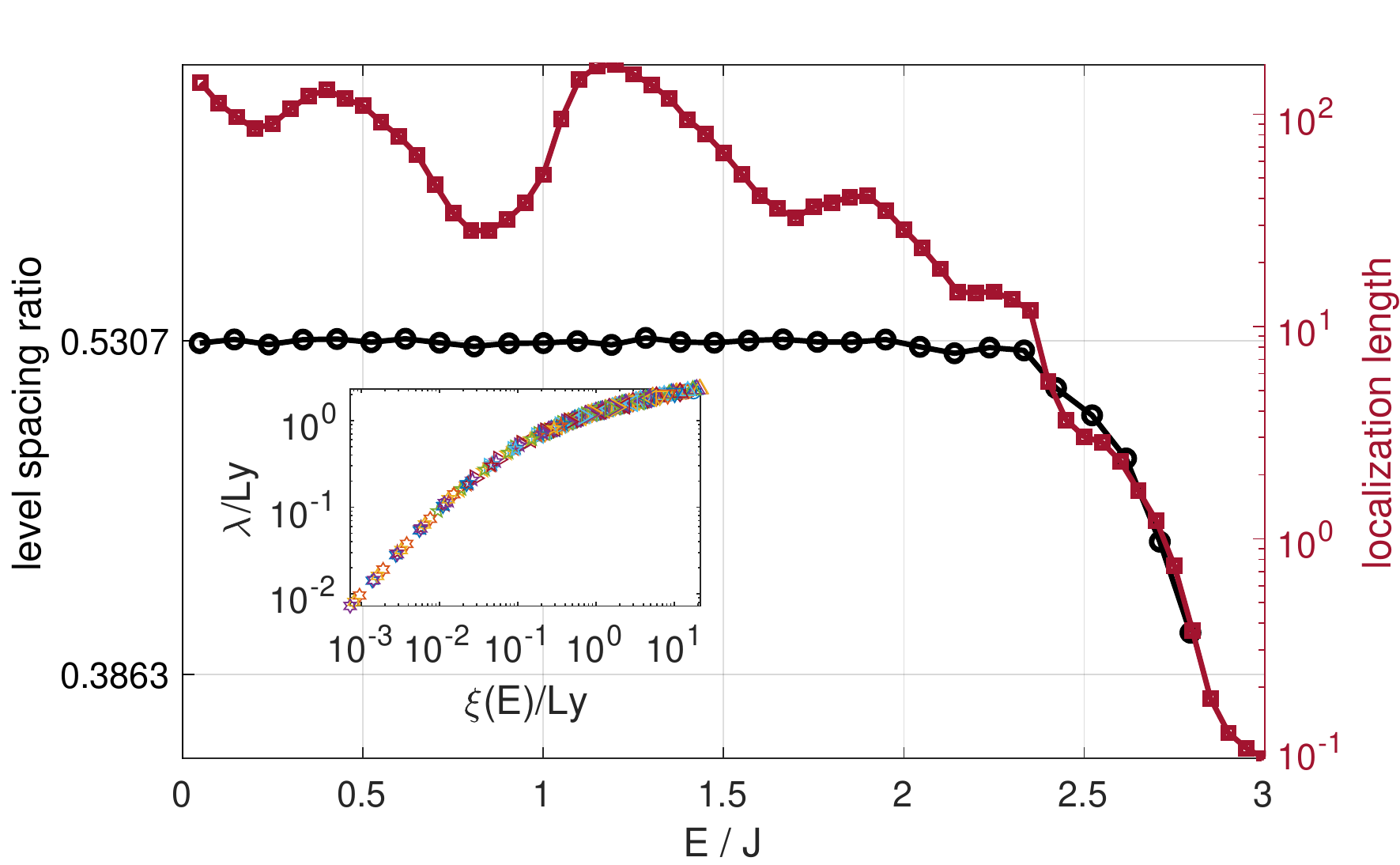} 
   \caption{Localization analysis for the time reversal symmetric Kitaev model without magnetic field $J_x=J_y=J_z=J, h=0$, which is mapped to a free Majorana fermion hopping on a honeycomb lattice with random $\pi$ flux disorder. Left axis: level spacing ratio calculated from 1000 disorder samples of system size $L_x=L_y=50$, where the Poisson value $2\ln2-1\approx 0.3863$ and the Gaussian orthogonal ensemble value $\approx 0.5307$ are indicated. Right axis: localization length calculated in quasi-1D stripes with $L_y=8,16,32,64,128$ while $L_x\leq 10^6$. The recursive iteration for Green's function is stopped when relative deviation of the smallest Lyapunov exponent gets smaller than 0.1 or $L_x=10^6$. Inset shows the data collapse of the one parameter scaling ansatz, for different stripe widths and energies.}
   \label{fig:masslessLocalization}
   \end{figure}

For comparison, we summarize the dynamical exponents fit from different physical observables and for different parameter regime as in Table.~\ref{table:dynamicalExponent}. 
\begin{table}[h]
\begin{center}
\begin{tabular}{c c c c c c c c}
\toprule
$J_z/J$ & $J_x/J$ & $J_y/J$ & $\tilde{h}$ & $z_{zz}$ & $z_{xx}$ & $z_{ent}$\\
\midrule
1.0 & 1.0 & 1.0 & 0.25 & 2.5(2) & 2.7(3) & 2.4(1) \\
2.0 & 0.5 & 0.5 & 0.25 & 2.6(2) & 3.4(2) & 2.6(1)\\
1.0 & 1.0 & 1.0 & 0.00 & 2.0(5) & 1.7(3) & 2.1(2)\\
\bottomrule
\end{tabular}
\end{center}
\caption{Fit dynamical exponent. 
}
\label{table:dynamicalExponent}
\end{table}%


\section{Fermionization for Hamiltonian and state}
Below we review two alternative fermionization approaches for the Kitaev model, and draw the connections between them. Generally speaking, Kitaev's original gauge theory approach is physically transparent and looks more symmetric in putting gauge field on every link, while the Jordan Wigner transformation
approach~\cite{Feng2006} is practically convenient without gauge redundancy. The latter can be achieved from the former by gauge fixing, with proper care taken for the boundary terms. 

\subsection{Jordan-Wigner transformation}
As a warmup let us first consider only one zig-zag row composed of $x$-links and $y$-links: 
$\sum_{j\in A}\sigma_j^x \sigma_{j+1}^x + \sum_{j\in B}\sigma_j^y \sigma_{j+1}^y$.
It can be readily solved by performing a zigzag Jordan-Wigner transformation:
\begin{equation}
\begin{aligned}
\sigma _{j\in  A}^x=\left(\prod _{i=1}^{j-1}\sigma _i^z\right)b_j,
\quad
\sigma _{j\in  A}^y=+\left(\prod _{i=1}^{j-1}\sigma _i^z\right)\alpha _j,
\quad
\sigma_{j\in A}^z=-i b_j\alpha _j,
\quad
\sigma _{j\in  B}^x=\left(\prod _{i=1}^{j-1}\sigma _i^z\right)\beta _j,
\quad
\sigma _{j\in  B}^y=-\left(\prod _{i=1}^{j-1}\sigma _i^z\right)b_j,
\quad
\sigma _{j\in B}^z=i \beta _jb_j,
\end{aligned}
\end{equation}
such that the Ising exchange interactions on x(y)-links are simplified to Majorana fermion hopping terms
\begin{equation}
\sigma _{j\in  A}^x\sigma _{j+1}^x=-i \alpha _j\beta _{j+1},
\qquad
\sigma _{j\in  B}^y\sigma _{j+1}^y=-i \alpha _{j+1}\beta _j.
\end{equation}
Now consider the honeycomb lattice, where spins are ordered row by row to be Jordan-Wigner transformed, and we label the sites by row and column indices $(i,j)$. The above intra-row interactions remain unchanged, while the inter-row couplings are simply 
\begin{equation}
\sigma _{i,j\in A}^z\sigma _{i+1,j}^z=-i \alpha _{i,j} \beta _{i+1,j}\left(i b_{i,j}b_{i+1,j}\right)
\equiv -i \alpha _{i,j} \beta _{i+1,j}u_{i,j}^z,
\end{equation}
where a static link variable $u_{i,j}^z=\pm1$ emerges on every $z$-link. Therefore the three anisotropic Ising interactions become free Majorana hopping terms coupled with classical $\mathbb{Z}_2$ variables. Since Jordan-Wigner transform is a non-local unitary, the boundary term in each row yields a non-local term:
\begin{equation}
\sigma _{i, 2L_x}^y\sigma _{i, 1}^y = - i\alpha _{i, 1}\beta _{i, 2L_x} \left(-\prod _{k=1}^{2L_x}\sigma _{i, k}^z\right)
\equiv i\alpha _{i, 1}\beta _{i, 2L_x} u_{i}^y,
\end{equation}
where $u_i^y$ commutes with the Hamiltonian therefore serves as a conserved quantum number, separating the fermion system into disconnected sectors with distinct boundary condition. In other words, the spin model with a definite boundary condition contain two fermion sectors with distinct boundary condition for fermions, which is reminiscent of the 1D quantum Ising critical point. 

Here we briefly comment that the boundary condition in fermion basis is relevant to the topological ground state degeneracy, which is generated by moving two $\mathbb{Z}_2$ local gauge fluxes around the torus before annihilation, equivalent to pumping a global $\mathbb{Z}_2$ flux for fermions~\cite{Nayak2007}. In the Abelian phase (toric code), one obtains 4-fold degeneracy by the freedom of pumping global $\mathbb{Z}_2$ flux along either direction. In the non-Abelian Ising phase, there are only 3-fold degeneracy,
because due to the Majorana zero mode trapped by the flux, pumping global flux along both directions would leave a global fermionic excitation behind and take the state away from ground state manifold. 

Under such Jordan Wigner transformation, the prequench initial spin polarized product state that satisfies
\begin{equation}
\sigma_{j\in A}^z |\Psi_0\rangle = - i b_{j\in A}\alpha _j |\Psi_0\rangle = - |\Psi_0\rangle ,
\qquad
\sigma_{j\in B}^z |\Psi_0\rangle = i \beta _j b_{j\in B} |\Psi_0\rangle =  |\Psi_0\rangle ,
\end{equation}
is transformed to a product state of local Majorana dimers satisfying $ i b_{j\in A}\alpha _j=1$, $ i \beta _j b_{j\in B}=1$. By grouping $(b,b)$ and $(\alpha,\beta)$, it is a maximally entangled EPR state between the local Majorana bilinears and the gauge field on $z$-link:
\begin{equation}
|\Psi_0\rangle = \prod_j \frac{1}{\sqrt{2}}
\left(|i\alpha_j\beta_{j-e_z}=1\rangle | u_j=1\rangle
+ |i\alpha_j\beta_j=-1\rangle | u_j=-1\rangle\right).
\end{equation}
Therefore the prequench state is a gauged fermion vacuum state with disordered gauge configurations. The post-quench Hamiltonian can be viewed as a Bogoliubov-de-Gennes superconducting Hamiltonian with gauged pairing and hopping terms, which would create fermion pairs from the gauged vacuum and allow them to propagate. 
On the other hand, such initial state projects onto the sector that locks opposite boundary condition between spin and fermion representations i.e. $-\sigma^y\sigma ^y = - i\alpha \beta $ on boundary while $\sigma^y\sigma ^y = - i\alpha \beta $ in the bulk. We'll work with the anti-periodic boundary condition for the spin model, and hence periodic boundary condition for the Majorana fermions.

\subsection{Redundant gauge theory}
Kitaev's original solution~\cite{Kitaev2005} is to extend the Hilbert space by rewriting each spin operator by four Majorana fermions with a physical constraint:
\begin{equation}
\sigma _j^{x(y)(z)}=i b_j^{x(y)(z)}c_j,
\qquad
-i \sigma _j^x\sigma _j^y\sigma _j^z=1=b_j^xb_j^yb_j^zc_j.
\end{equation}
Then the spin Hamiltonian rewritten in terms of Majorana fermions are always trivially conserved by $e^{i\pi G}\equiv b_j^xb_j^yb_j^zc_j$. In other words, the subsequent Hamiltonian in terms of Majorana fermions by definition has a $\mathbb{Z}_2$ gauge symmetry, generated by $e^{i\pi G}$. Notice that $G$ is the $\mathbb{Z}_2$ analogue of the Gauss law operator $n_j-\sum _k E_{j,k}$ as in the conventional U(1) gauge theory. The constraint $b_j^xb_j^yb_j^zc_j=1$ actually picks up the gauge neutral sector among all other super-selection sectors to be the physical space, as a superposition of all gauge equivalent configurations. 

In the exact solvable regime, the $\mathbb{Z}_2$ gauge field on every link $u_j^{x(y)(z)}\equiv i b_j^{x(y)(z)}b_{j-e_{x(y)(z)}}^{x(y)(z)}$ is conserved. To
avoid extensive redundancy, one can always use the gauge transformation to fix the gauge such that $u_j^{x(y)}=1$ on every link in the bulk. In other words,
the gauge field is oriented only towards z-direction, analogous to the Landau gauge. However, one should pay special attention when the lattice is
not on a planar but a compact manifold such as torus. In a torus, there could be global fluxes which can neither be detected locally nor be locally
gauged away. Therefore, there must remain $u_j^{x(y)}\neq 1$ on certain boundary links that account for the global fluxes. One can also see this by
taking the product of gauge neutral constraints along one zig-zag row: 
$\prod _{j\in \text{row}}u_j^xu_j^y=-\prod _{j\in\text{row}}\sigma _{j}^z$.
Therefore one may fix $u_j^{x(y)}=1$ in the bulk but must allow one boundary x(y)-link on every row fluctuating, which is equal to $-\prod _{j\in\text{row}}\sigma _{j}^z$. In this way, we reach exactly the same fermionic Hamiltonian with a nonlocal boundary term, as derived from Jordan-Wigner transformation. And in this way we shall see that the $\mathbb{Z}_2$ variable emergent in Jordan Wigner transformation has the physical meaning of a gauge field. 

Last but not least, one should notice that the onset of magnetic field does not explicitly spoils the $\mathbb{Z}_2$ gauge symmetry of fermion since that is just a gauge redundancy in the extended Hilbert space. Instead, the magnetic field breaks another set of local $\mathbb{Z}_2$ symmetries which stand for the gauge flux conservation. 

\section{Observables at exact solvable regime}

For convenience let us label the unitcell($z$-link) by $j$ in the following and group the Majorana fermions into a $2N$-dimensional vector
$\zeta \equiv \left(
\begin{array}{c}
 \alpha  \\
 \beta  \\
\end{array}
\right)$,
such that $\zeta _{j=1,\ldots ,N}$ specifies $\alpha $, and $\zeta _{j=N+1,\ldots ,2N}$ specifies $\beta $. The Gaussian evolution operator acting on the free Majorana fermion reduces to an 2$N$-by-2$N$ evolution matrix 
$\zeta (t)= e^{\left[i t \hat{H}, \right]}\zeta =e^{-i 2t H}\zeta$.
Therefore one can keep track of the time evolved two-Majorana-fermion covariant matrix:
\begin{equation}
\Gamma (t)\equiv \frac{i}{2} \langle [\zeta (t) ,\zeta ^T (t)]\rangle =e^{-i 2t H}i\tau^y\otimes (\oplus_ju_j) e^{i 2t H},
\end{equation}
which is skew-symmetric matrix. The matrix elements immediately give the spin dimer expectations expressed in terms of gauged Majorana fermion bilinears:
\begin{equation}
\begin{aligned}
\langle D_j^z(t) \rangle = \langle i u_j\alpha _j(t)\beta _j(t) \rangle =u_j\Gamma (t)_{j,j+N}, 
\qquad 
\langle D_j^x(t) \rangle =\langle i \alpha _j(t)\beta _{j+n_2}(t) \rangle =\Gamma (t)_{j,j+n_2+N}.
\end{aligned}
\end{equation}
The disconnected dimer correlation functions expressed in terms of the four-Majorana-fermion correlation functions can be decomposed using Wick's
theorem
\begin{equation}
\begin{aligned}
&C_j^{zz}(r,t) \equiv \langle D_j^z(t)D_{j+r}^z(t)\rangle = u_ju_{j+r}\langle i\alpha_j(t)\beta _j(t)i\alpha_{j+r n_1}(t)\beta _{j+r
n_1}(t)\rangle \\
& = u_ju_{j+r n_1}\left\{ \langle i\alpha_j(t)\beta _j(t)\rangle \langle i\alpha_{j+r n_1}(t)\beta _{j+r n_1}(t)\rangle
-\langle i\alpha_j(t)\alpha _{j+r n_1}(t)\rangle \langle i\beta_j(t)\beta _{j+r n_1}(t)\rangle -\langle
i\alpha_j(t)\beta _{j+r n_1}(t)\rangle \langle i\alpha_{j+r n_1}(t)\beta _j(t)\rangle \right\}\\
& = u_ju_{j+r n_1} \left( \Gamma(t)_{j, j + N}  \Gamma(t)_{j+r n_1 , j+r n_1 + N} - \Gamma(t)_{j, j + r n_1}  \Gamma(t)_{j +N, j+r n_1 + N} - \Gamma(t)_{j, j + r n_1 + N}  \Gamma(t)_{j+r n_1 , j + N} \right).
\end{aligned}
\end{equation}
Likewise for $C^{xx}$. 

In each gauge configuration, the time evolved density operator is Gaussian. Therefore the reduced density operator of the half-partitioned fermion system is also a Gaussian density operator that can be
disentangled into reduced canonical fermions denoted as $d_n$, $n=1,\cdots,N/2$:
\begin{equation}
\hat{\rho }_r \equiv \frac{e^{\frac{-1}{4}\zeta^T  H_{\text{ent}}\zeta }} 
{ \text{Tr}\left(e^{\frac{-1}{4}\zeta^T H_{\text{ent}}\zeta }\right)}
=\prod _n\frac{e^{-\xi _nd_n^{\dagger
}d_n}}{1+e^{-\xi _n}}=\prod _n\left(\frac{1}{1+e^{\xi _n}}d_n^{\dagger }d_n+\frac{1}{1+e^{-\xi _n}}d_nd_n^{\dagger }\right).
\end{equation}
This Gaussian density operator is faithfully encoded in the covariant matrix
\begin{equation}
\text{Tr} \left(\hat{\rho}_r i\zeta\zeta^T \right)
=2i\left(1+e^{- H_{\text{ent}}}\right)^{-1}
=\Gamma (t)_{\text{subblock}} + i ,
\end{equation}
which is just the subsystem block of the total covariant matrix $\Gamma(t)$. Therefore one can extract the eigenvalue of the disentangled density operator (occupation of the reduced canonical fermion mode) and the entanglement entropy:
\begin{equation}
\rho _n(t)=\frac{1}{1+e^{\xi _n}}=\text{spec}\left\{\frac{1-i\Gamma (t)_{\text{subblock}}}{2}\right\},
\qquad
S_v(t)=-\sum _{n=1}^{N/2} \left( \rho _n\ln  \rho _n+\left(1-\rho _n\right)\ln  \left(1-\rho _n\right) \right) \leq  N ln2/2.
\end{equation}
The upper-bound is reached if and only if each canonical fermion is maximally entangled with a flat spectrum for the subblock of covariant matrix i.e. $\rho_n=1/2$.

\section{Turning on Majorana interaction}
\subsection{Resonant Majorana fermion interaction}

The non-interacting part of the Majorana fermion Hamiltonian written in terms of spinor of local Majorana fermions can be canonical transformed to\cite{Kitaev2005}:
\begin{equation}
\hat{H}=\frac{1}{2} \left(
\begin{array}{cc}
 \alpha  & \beta  \\
\end{array}
\right)H\left(
\begin{array}{c}
 \alpha  \\
 \beta  \\
\end{array}
\right)\\
\\
=\frac{1}{2}\left(
\begin{array}{cc}
 \gamma' & \gamma'' \\
\end{array}
\right) \tau^y\otimes \epsilon \left(
\begin{array}{c}
 \gamma' \\
 \gamma'' \\
\end{array}
\right)\\
\\
=\left(
\begin{array}{cc}
 c^{\dagger } & c \\
\end{array}
\right) \tau^z\otimes \epsilon \left(
\begin{array}{c}
 c \\
 c^{\dagger } \\
\end{array}
\right),
\end{equation}
where $H=-H^T=-H^*$, $\tau^\mu$ is the pauli matrix acting on the Nambu particle-hole spinor space, and $\epsilon$ denotes the diagonal matrix with single fermion eigenstate energy as entries. The canonical Majorana fermions $\gamma$ (complex Bogoliubov fermions $c$ ) are related to the local fermions by orthogonal(unitary) matrix:

\begin{equation}
\left(
\begin{array}{c}
 \alpha  \\
 \beta  \\
\end{array}
\right)=Q \left(
\begin{array}{c}
 \gamma' \\
 \gamma'' \\
\end{array}
\right)=\sqrt{2}U \left(
\begin{array}{c}
 c \\
 c^{\dagger } \\
\end{array}
\right),
\qquad
Q=\sqrt{2}\left(
\begin{array}{cc}
 \text{Re}\left(\psi _A\right) & \text{Im}\left(\psi _A\right) \\
 -\text{Im}\left(\psi _B\right) & \text{Re}\left(\psi _B\right) \\
\end{array}
\right),\qquad
U=\left(
\begin{array}{cc}
 -i \psi _A & i \psi _A^* \\
 \psi _B & \psi _B^* \\
\end{array}
\right),
\end{equation}
that satisfy
\begin{equation}
U^{\dagger } H U=\left(
\begin{array}{cc}
 \epsilon  & 0 \\
 0 & -\epsilon  \\
\end{array}
\right)\equiv \tau^z\otimes\epsilon,\qquad
Q^T H Q=\left(
\begin{array}{cc}
 0 & -i \epsilon  \\
 i \epsilon  & 0 \\
\end{array}
\right)\equiv \tau^y\otimes \epsilon.
\end{equation}
Now we bring the canonical transformation to the Majorana Hubbard interaction $ \tilde{h} \sum_{\text{Y}} u_{i,j}u_{i,k}u_{i,l} \left(\beta_i\alpha_j\alpha_k\alpha_l - \alpha_i\beta_j\beta_k\beta_l\right)$. For convenience here we use $i,j,k,l$ to label the spin site instead of unit-cell. Consider two inversion related Y junctions tied to the same $z$-link, with site labeled in counter-clockwise ordering,
\begin{equation}
\begin{aligned}
\beta _1\alpha _1\alpha _2\alpha _3
&=4\sum _{a,b,c,d}\left(\psi _{B,1,a}c_a+h.c.\right)\left(-i\psi_{A,1,b}c_b+h.c.\right)\left(-i\psi_{A,2,c}c_c+h.c.\right)\left(-i\psi_{A,3,d}c_d+h.c.\right)\\
&=4\sum _{m,n}\left(\text{Re}\left(\psi _{B,1,m}^*\psi _{A,1,m}\right)\text{Im}\left(\psi _{A,2,n}^*\psi _{A,3,n}\right)+\text{cycl}.\text{perm}\right)\left(2c_m^{\dagger
}c_m-1\right)\left(2c_n^{\dagger }c_n-1\right)+\cdots\\
&=-4\sum _{m,n}\left(\text{Re}\left(\psi _{B,1,m}^*\psi _{A,1,m}\right)\text{Im}\left(\psi _{A,2,n}^*\psi _{A,3,n}\right)+\text{cycl}.\text{perm}\right)\gamma
_m'\gamma _m''\gamma _n'\gamma _n''+\cdots
\end{aligned}
\end{equation}
\begin{equation}
\begin{aligned}
-\alpha _1\beta _1\beta _2\beta _3 &=-4\sum _{a,b,c,d}\left(-i\psi_{A,1,a}c_a+h.c.\right)\left(\psi _{B,1,b}c_b+h.c.\right)\left(\psi _{B,2,c}c_c+h.c.\right)\left(\psi _{B,3,d}c_d+h.c.\right)\\
&=4\sum _{m,n}\left(\text{Re}\left(\psi _{A,1,m}^*\psi _{B,1,m}\right)\text{Im}\left(\psi _{B,2,n}^*\psi _{B,3,n}\right)+\text{cycl}.\text{perm}\right)\left(2c_m^{\dagger
}c_m-1\right)\left(2c_n^{\dagger }c_n-1\right)+\cdots\\
&=-4\sum _{m,n}\left(\text{Re}\left(\psi _{A,1,m}^*\psi _{B,1,m}\right)\text{Im}\left(\psi _{B,2,n}^*\psi _{B,3,n}\right)+\text{cycl}.\text{perm}\right)\gamma
_m'\gamma _m''\gamma _n'\gamma _n''+\cdots
\end{aligned}
\end{equation}
where cycl.perm denotes the cyclic permutation counterparts. One could qualitatively verify the above result by the fact that the chiral three spin interaction $\sigma_{A(B)}^x\sigma_{A(B)}^y\sigma_{A(B)}^z$ respects three-fold rotation symmetry $\mathcal{C}_3$ but behaves odd under either time reversal or mirror reflection symmetry 
\begin{equation}
\begin{aligned}
&\mathcal{C}_3: \sigma^{x/y/z} \to \sigma^{y/z/x},\ (u, \alpha_{1/2/3}, \beta_{1/2/3})\to(u,\alpha_{2/3/1},\beta_{2/3/1}),\\
&\mathcal{T}: \sigma^\mu \to -\sigma^\mu,\ (u, \alpha, \beta)\to(u,\alpha,-\beta),
\quad
\mathcal{M}_z: \sigma_{A/B}^{x/y/z} \to -\sigma_{B/A}^{y/x/z},\ (u, \alpha, \beta)\to(u,\beta,-\alpha).
\end{aligned}
\end{equation}
And $\mathcal{M}_z$ maps $\beta_1\alpha_1\alpha_2\alpha_3$ to $\alpha _1\beta _1\beta _2\beta _3$, which enforces the sign difference. Inside the weight, taking the imaginary part of the wave-function overlap between the same sublattice enforces the mirror reflection odd condition. 
These diagonal part contributes to the leading order resonant interaction. Generally, the omitted off-diagonal terms can be Schrieffer-Wolff rotated to yield higher order correction to the resonant interaction, which usually only plays a role in exponential longer time. Here we just take the leading order term with the gauge coupling to get the symmetric interaction coupling matrix:
\begin{equation}
V_{m,n}
=8\tilde{h}\sum _iu_{i,j}u_{i,k}u_{i,l}\text{Re}\left(\psi _{i,m}^*\psi _{j,m}\right)\text{Im}\left(\psi _{k,n}^*\psi _{l,n}\right)+\text{cycl}.\text{perm}+(m\leftrightarrow n),
\end{equation}
in which $j,k,l$ are nearest neighbours arranged in counter-clockwise order surrounding site $i$, and $i$ is summed over both $A$ and $B$ sublattices. In this way, we arrive at our final effective Hamiltonian to leading order:
\begin{equation}
\label{eq:Hgamma}
\hat{H}_\gamma=-\sum_{n=1}^{N}\epsilon_n i\gamma'_n\gamma''_n - \frac{1}{4}\sum_{m,n=1}^{N} V_{m,n} \gamma'_m\gamma''_m\gamma'_n\gamma''_n+\cdots.
\end{equation}

For completeness we here show the conserved canonical fermion parity in each random gauge configuration $\langle i\gamma_n'\gamma_n'' \rangle$, and the corresponding mean-field shift of energy due to the resonant interaction, see Fig.~\ref{fig:canonicalFermionParity}. 

\begin{figure}[h] 
   \centering
   \includegraphics[width=8cm]{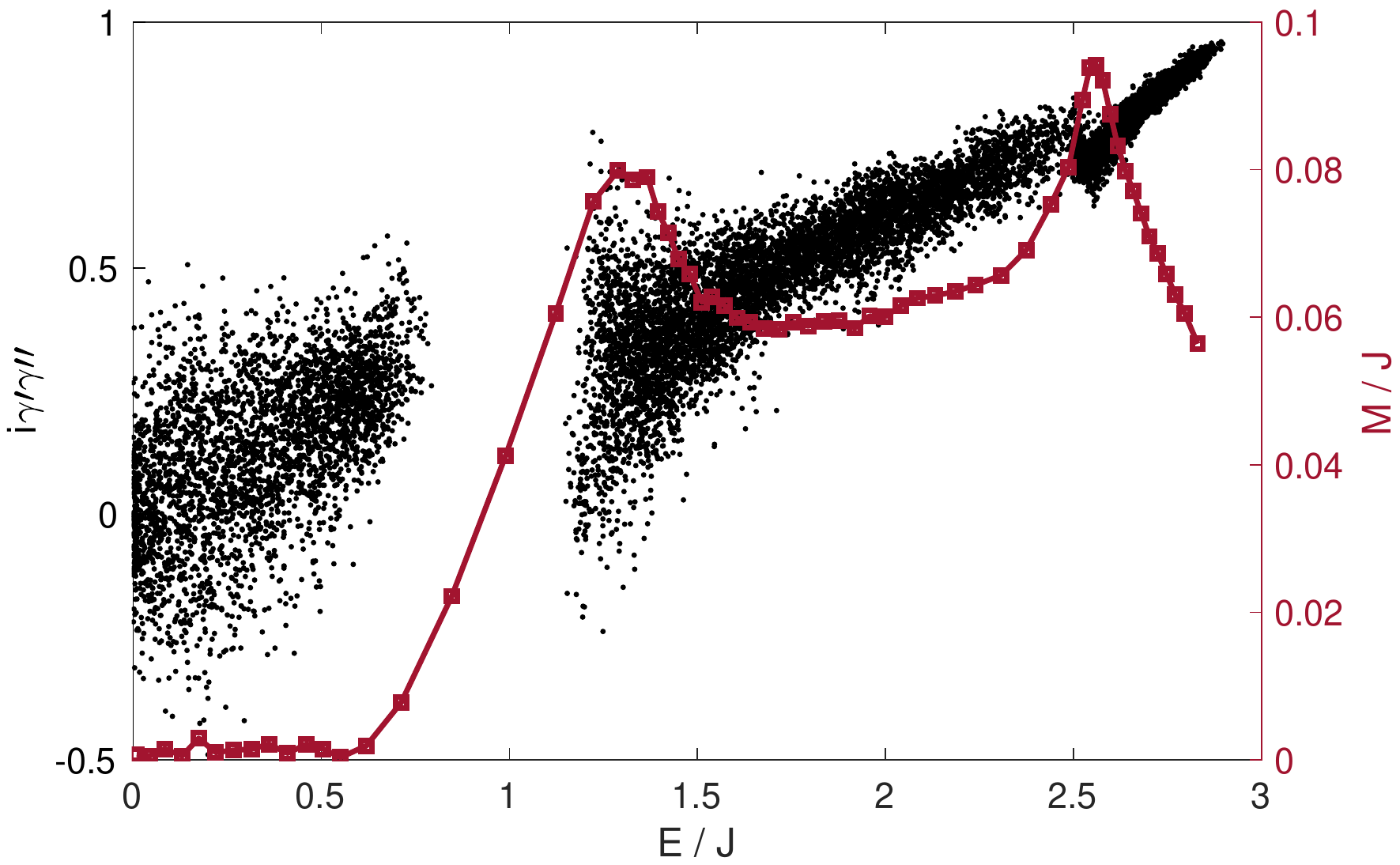} 
   \caption{Left axis: $\langle i\gamma_n'\gamma_n'' \rangle$. Each dot is associated with one disordered gauge configuration. The distribution lies between the infinite temperature where $\langle i\gamma'\gamma'' \rangle=0$ and zero temperature where $\langle i\gamma'\gamma'' \rangle=1$. Right axis: mean-field energy shift $M_n \equiv \sum_m V_{m,n}\langle i\gamma_m'\gamma_m''\rangle /4$. Parameters are $J_x=J_y=J_z=J, \tilde{h}=0.25J, L_x=16, L_y=4$, 200 disorder samples. }
   \label{fig:canonicalFermionParity}
\end{figure}

\subsection{Statistics of omitted off-diagonal interaction}

Here we do a statistical analysis for the off-diagonal interaction terms that were omitted. For simplicity let's fix the gauge such that nonzero gauge connection occurs only at $z$-link, and take a unit-cell and label the
6 sites involved as in inset of Fig.~\ref{fig:offDiagIntAnalysis}. The Majorana Hubbard interaction is the product of certain Majorana fermion bilinears on bonds $\hat{V}=\tilde{h}\sum
_{\text{unit}-\text{cell}}u_0\left(i\alpha _0\beta _1i\beta_2\beta _3+i\alpha _0\beta _1i\alpha _4\alpha _5\right)$,
which can be straightforwardly verified by $\sigma _2^x\sigma _1^z\sigma _3^y=\left(\sigma _0^z\sigma _1^z\right)\left(\sigma _2^x\sigma _0^z\sigma
_3^y\right)=\left(-u_0i\alpha _0\beta _1\right)\left(i\beta_3\beta _2\right)$, likewise for $\sigma _4^x\sigma _0^z\sigma _5^y$.
Using $\alpha _j=\sqrt{2}\sum _n\left(-i \psi _{j,n} c_n+h.c.\right), \beta _j=\sqrt{2}\sum _n\left(\psi _{j,n} c_{j,n}+h.c.\right)$, the involved
fermion bilinears can be decomposed into off-diagonal and diagonal parts:
\begin{equation}
\begin{split}
&i\alpha _0\beta _1=\sum _{m\neq n}2i \left(-i \psi _{0,m}c_m+h.c.\right)\left(\psi _{1,n}c_n+h.c.\right)+\sum _p2\text{Re}\left(\psi _{0,p}\psi_{1,p}^*\right)i\gamma_p'\gamma _p'',\\
&i\beta_2\beta _3=\sum _{m\neq n}2i \left( \psi _{2,m}c_m+h.c.\right)\left(\psi _{3,n}c_n+h.c.\right)+\sum _p2\text{Im}\left(\psi _{2,p}^*\psi
_{3,p}\right)i\gamma_p'\gamma _p'',\\
&i\alpha _4\alpha _5=\sum _{m\neq n}2i \left(-i \psi _{4,m}c_m+h.c.\right)\left(-i\psi_{5,n}c_n+h.c.\right)+\sum _p2\text{Im}\left(\psi
_{4,p}^*\psi _{5,p}\right)i\gamma_p'\gamma _p''.
\end{split}
\end{equation}
A rough overview: the product of two diagonal fermion bilinears $\gamma _p'\gamma _p''\gamma _q'\gamma _q''$ contributes to the l-bit type
resonant interaction terms that have been shown previously. The product of one diagonal bilinear $\gamma _p'\gamma _p''$ and one off-diagonal
bilinear $\gamma _m\gamma _n$ contributes to the off-diagonal interaction term that flips the fermion parity of two canonical fermions, which can
be further decomposed into an assistant pairing term $\sim \left(1-2c_p^{\dagger }c_p\right)c_m^{\dagger }c_n^{\dagger }$ and an assistant hopping
term $\sim \left(1-2c_p^{\dagger }c_p\right)c_m^{\dagger }c_n$, mediated by the presence or absence of other canonical fermion modes. Since the
density term depending on index $p$ is to be integrated out, the assistant pairing/hopping interaction is roughly proportional to the overlap of canonical fermion
wave-function of $m$ and $n$ on the same given unit-cell. In contrast, the product of two off-diagonal bilinears $\gamma _m\gamma _n\gamma _p\gamma
_q(m\neq n\neq p\neq q)$ that changes the fermion parity of four canonical fermion modes requires the overlap of four localized fermion wave-function
on the same unit-cell, which are less dominant. In the following we do statistical analysis for the assistant hopping and pairing terms. 

\begin{figure}[h] 
   \centering
   \includegraphics[width=8cm]{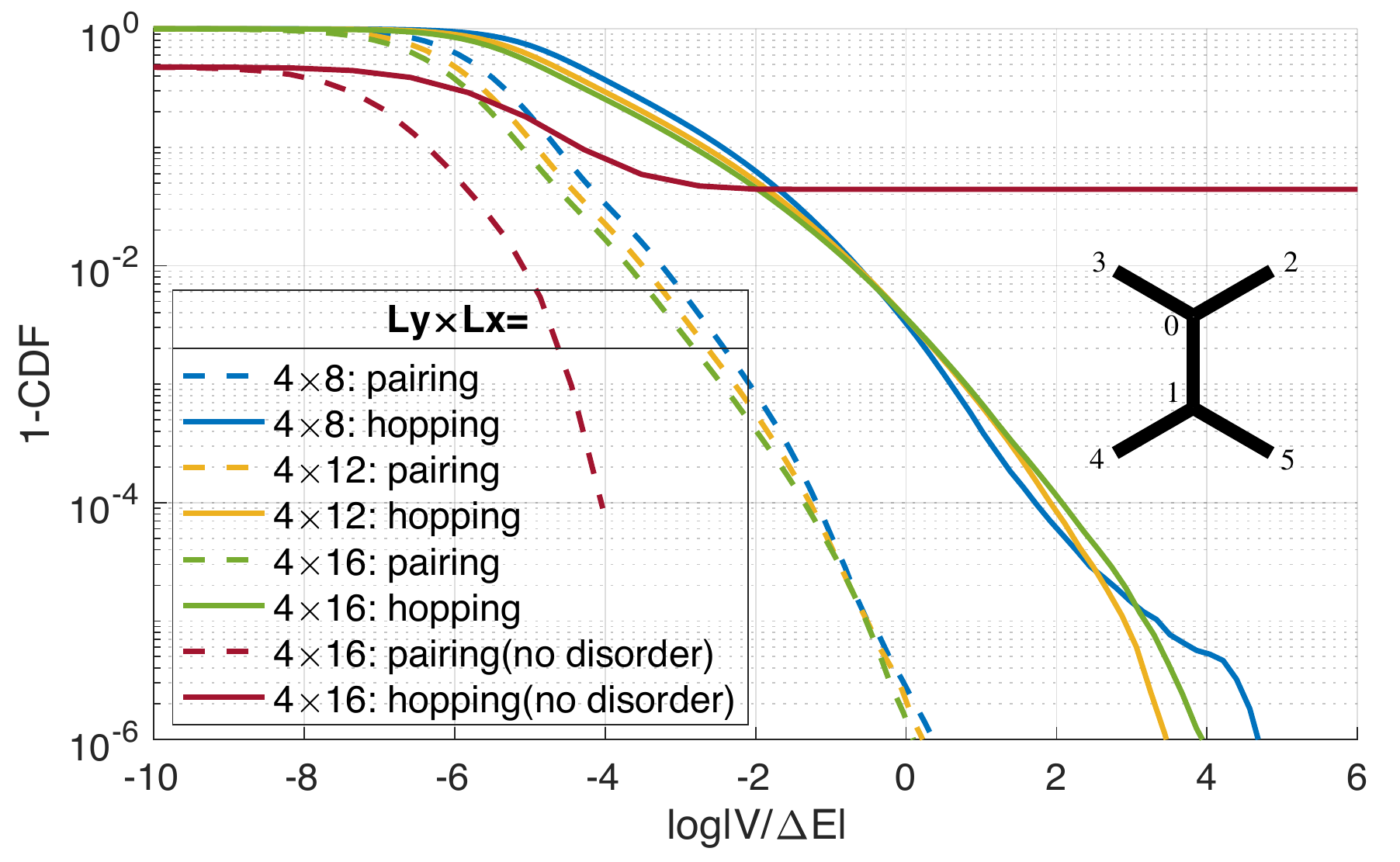} 
   \caption{Cumulative probability distribution function(CDF) for the off-diagonal interaction matrix element divided by the corresponding energy difference, in the eigenstate basis of exact solvable Hamiltonian in random gauge sectors. Varying system sizes are shown, as well as the assistant hopping and pairing types. The assistant hopping term is relatively easier to reach resonance than pairing term, with probability $\lesssim 0.5\%$. In comparison, in the zero flux sector without effective gauge flux disorder, the assistant hopping interaction shows strongly resonant contribution with $\ln|V/\Delta E| > 10^6$ with probability $\gtrsim 4\%$ among the same randomly sampled eigenstates. Therefore it is the disorder that suppresses the strong resonance. Parameters: $J_x=J_y=J_z=J, \tilde{h}=0.25J$, sampled over 200 disorder realization of random gauge configurations, with 100 randomly sampled eigenstate in each configuration. }
   \label{fig:offDiagIntAnalysis}
\end{figure}
In detail, for the Kitaev Hamiltonian at solvable point, we sample 200 random gauge configurations $\left\{u_j\right\}$ where $j$ labels the sites. For
each gauge configuration we randomly sample 100 eigenstates, which are determined by the canonical fermion pairity array $\left\{\nu _n\right\}$,
with $\nu _n\equiv i\gamma_n'\gamma _n''=\pm 1$being randomly sampled. We consider the assistant hopping(pairing) interactions
connecting these random states to the target states with canonical fermion modes $m,n$ being flipped: $\langle \overset{\rightarrow}{\nu }-2\nu _m\mp
2\nu _n|\hat{V}|\overset{\rightarrow}{\nu }\rangle$, for which the energy difference is $2\left(\epsilon _m\pm \epsilon _n\right)$. It can be shown that
for a given choice of $m,n$, 
\begin{equation}
\langle \overset{\rightarrow}{\nu }-2\nu _m\mp 2\nu _n |\hat{V}|\overset{\rightarrow}{\nu } \rangle =\tilde{h}\sum _p\nu _pF_{m,n,p}^{\pm },
\end{equation}
where each other inert canonical fermion mode contribute to the matrix element by a three-rank form factor tensor given by the wave-function overlap:
\begin{equation}
\begin{split}
F_{m,n,p}^+&\equiv 4\sum _{\text{unit}-\text{cell}}
\left(-i u_0\text{Re}\left(\psi _{0,p}\psi _{1,p}^*\right)\right)\psi _{4,m}^*\psi _{5,n}^*-u_0\text{Im}\left(\psi
_{4,p}^*\psi _{5,p}\right)\psi _{0,m}^*\psi _{1,n}^*+(0\rightarrow4\rightarrow5\rightarrow0)\\
&+i u_0\text{Re}\left(\psi _{0,p}\psi _{1,p}^*\right) \psi _{2,m}^*\psi _{3,n}^*-u_0\text{Im}\left(\psi
_{2,p}^*\psi _{3,p}\right)\psi _{0,m}^*\psi _{1,n}^*+(1\rightarrow2\rightarrow3\rightarrow1)
-(m\leftrightarrow n),\\
F_{m,n,p}^-&\equiv 4\sum _{\text{unit}-\text{cell}}
i u_0\text{Re}\left(\psi _{0,p}\psi _{1,p}^*\right)\psi _{4,m}^*\psi _{5,n}-u_0\text{Im}\left(\psi _{4,p}^*\psi
_{5,p}\right)\psi _{0,m}^*\psi _{1,n}+(0\rightarrow4\rightarrow5\rightarrow0)\\
&+i u_0\text{Re}\left(\psi _{0,p}\psi _{1,p}^*\right) \psi _{2,m}^*\psi _{3,n}-u_0\text{Im}\left(\psi
_{2,p}^*\psi _{3,p}\right)\psi _{0,m}^*\psi _{1,n}+(1\rightarrow2\rightarrow3\rightarrow1)
-(m\leftrightarrow n),
\end{split}
\end{equation}
where $(0\rightarrow4\rightarrow5\rightarrow0)$ and $(1\rightarrow2\rightarrow3\rightarrow1)$ denotes the $C_3$ cyclic permutation counterparts, and $(m\leftrightarrow n)$ means transposing
m and n indices to enforce the anti-symmetric condition. The off-diagonal interaction matrix element divided by the energy difference serves as the
strength of the first order perturbative Schrieffer-Wolff transformation generator, and therefore determines the quality of our zeroth order perturbation
theory. Obtaining the interaction matrix elements for all possible $m,n=1,\ldots ,N$, we calculate the cumulative probability distribution function
of the logarithm of this dimensionless quantity $\ln \left|\frac{\langle \overset{\rightarrow}{\nu }'|\hat{V}|\overset{\rightarrow}{\nu } \rangle }{\Delta
E}\right|$, as shown in Fig.~\ref{fig:offDiagIntAnalysis}. The resonance condition with $\ln|V/\Delta E| \gtrsim 0$ occurs with probability $\lesssim 0.5\%$, which is the probability when our zeroth order perturbation theory would fail.


\subsection{Correlations with dephasing interaction}

In the presence of dephasing interaction, we need to transform the physical observables into the canonical fermion basis:
\begin{equation}
\langle D^z(t) \rangle \equiv \frac{1}{N2^N} \sum_{\{u\}}\sum_j
iu_j \langle \alpha_j(t)\beta_j(t)\rangle
=
\frac{1}{2^N}\sum_{\{u\}}\left(\frac{1}{N} \sum_j
u_j Q_{j,m}Q_{j,n}\right)\langle i\gamma_m(t)\gamma_n(t)\rangle.
\end{equation}
The two-dimer disconnected correlation function is equivalent to the four-Majorana correlation functions as follows:
\begin{equation}
\begin{aligned}
&C^{zz}(r,t)\equiv 
\frac{1}{N}\sum _j\langle D_j^z(t)D_{j+r}^z(t)\rangle 
=-\frac{1}{N2^N}\sum_{\{u\}}\sum _j
u_ju_{j+r n_1}\langle \alpha _j(t) \beta _j(t) \alpha _{j+r n_1}(t) \beta
_{j+r n_1}(t) \rangle \\
=&-\frac{1}{N2^N}\sum_{\{u\}}\sum _j \sum _{m\neq n\neq p\neq q =1}^{2N} u_ju_{j+r n_1}Q_{j,m}Q_{j+N,n}Q_{j+r n_1,p}Q_{j+r n_1+N,q}\langle \gamma _m(t)\gamma _n(t)\gamma _p(t)\gamma
_q(t)\rangle \\
=&\frac{1}{2^N}\sum_{\{u\}}\sum _{m\neq n\neq p\neq q =1}^{2N} \left(-\frac{1}{N}\sum _ju_ju_{j+r n_1}Q_{j,m}Q_{j+N,n}Q_{j+r n_1,p}Q_{j+r n_1+N,q}\right)\langle \gamma _m(t)\gamma _n(t)\gamma
_p(t)\gamma _q(t)\rangle \\
\equiv &\frac{1}{2^N}\sum_{\{u\}}\sum _{m\neq n\neq p\neq q =1}^{2N}  f_{m,n,p,q}^{zz}\langle \gamma _m(t)\gamma _n(t)\gamma _p(t)\gamma _q(t)\rangle
= \frac{1}{2^N}\sum_{\{u\}}\sum _{m < n < p < q =1}^{2N}  24A[f_{m,n,p,q}^{zz}] \langle \gamma _m(t)\gamma _n(t)\gamma _p(t)\gamma _q(t)\rangle.
\end{aligned}
\end{equation}
Notice that due to the orthogonality of $Q$ matrix $\sum_mQ_{j,m}Q_{i,m}=\delta_{i,j}$, the terms with overlapping indices drops out such that only off-diagonal Majorana correlation functions with $m\neq n\neq p \neq q$ contribute. $A[f_{m,n,p,q}^{zz}]$ means anti-symmetrizing the form factor $f^{zz}$ with respect to indices $m,n,p,q$, using the anticommutation relation of Majorana fermions. Likewise for $\langle D^x\rangle$ and $C^{xx}$. 

Unlike the free fermion case, the four-point Majorana correlation function cannot be immediately factorized by Wick's theorem into simple product of two-point Majorana correlation functions. Each canonical Majorana fermion doublet $\gamma_n \equiv (\gamma_n', \gamma_n'')^T$ is evolved effectively by a particle-hole superposition of Gaussian operators:
\begin{equation}
\gamma _n(t)=e^{i t\left[H_{\text{eff}},\right]}\gamma _n=\sum _{\nu =\pm 1}e^{-i 2t \nu  \epsilon _n}e^{\frac{1}{4}\gamma ^T \nu A_n\gamma }\frac{1+\nu
 \tau ^y}{2}\gamma _n,
 \qquad
A_n(t)\equiv -i 2t \left(\tau ^y\otimes (\oplus _jV_{n,j})\right).
\end{equation}
Then the two-point Majorana fermion correlation functions can be expressed as
\begin{equation}
\langle \gamma _m(t)\gamma _n(t)^T\rangle =\sum _{\mu , \nu =\pm }e^{-i 2t\left(\mu  \epsilon _m+\nu  \epsilon _n\right)} \frac{1+\mu
 \tau ^y}{2}e^{-i 2t \nu  V_{n,m}\tau ^y}\langle e^{\frac{1}{4}\gamma ^T \left(\mu  A_m+\nu  A_n\right)\gamma } \gamma _m \gamma _n^T\rangle
\frac{1-\nu  \tau ^y}{2}.
\end{equation}
The key is to evaluate the expectation of $e^{\frac{1}{4}\gamma  \left(\mu  A_m+\nu  A_n\right)\gamma } \gamma _m \gamma _n^T$ over the initial
state $|\psi _{\{u\}}\rangle$. This can be generally done because the initial density matrix and the exponential operator are both Gaussian. For simplification in a given gauge configuration, we have the $A$ sublattice absorb the neighboring gauge field on $z$-link $\alpha _j\to u_{j}\alpha _j$ such that initial state is rotated to a clean Fock vacuum, while the gauge field dependence is absorbed by $Q\to (u, 1)Q$ matrix. Turning back to the original fermion basis $\gamma ^T A \gamma =\zeta ^T Q A Q^T\zeta$ where $\mu  A_m+\nu
 A_n\equiv A$ and $\zeta_j$ labels the Majorana doublet in unit-cell $j$, we have
\begin{equation}
\langle \psi _{\{u\}}|e^{\frac{1}{4}\gamma ^TA \gamma } \gamma _m \gamma _n^T|\psi _{\{u\}}\rangle 
= \sum_{i,j}
Q_{i,m}\langle \text{vac}|e^{\frac{1}{4}\zeta
^TQ A Q^T \zeta } \zeta _i \zeta _j^T|\text{vac}\rangle Q_{j,n},
\end{equation}
where $i \alpha_j \beta_j |\text{vac}\rangle = |\text{vac}\rangle$, and every eigenvector matrix $Q$ has been regauged accordingly. The quantity in the middle is a vacuum expectation of the product of Gaussian operator and Majorana fermions, which can be evaluated using the generic
formula that is to be derived later. 

 Likewise, the four-Majorana-fermion disconnected correlation function can be reduced to the expectation of the product of Gaussian operator and
four Majorana fermions:

\begin{equation}
\begin{aligned}
&\langle \gamma _m(t)\gamma _n(t)\gamma _p(t)\gamma _q(t)\rangle 
=\sum _{\mu , \nu , \kappa , \lambda =\pm 1}\sum_{m',n',p',q'=1}^{N}
e^{-i 2\left(\mu  \epsilon
_m+\nu  \epsilon _n+\kappa  \epsilon _p+\lambda  \epsilon _q\right)t} \\
\times&\left(\frac{1+\mu  \tau ^y}{2}e^{-i t 2\left(\nu  V_{n,m}+\kappa  V_{p,m}+\lambda
 V_{q,m}\right)\tau ^y}\right)_{m,m'}\left(\frac{1+\nu  \tau ^y}{2}e^{-i t 2\left(\kappa  V_{p,n}+\lambda  V_{q,n}\right)\tau ^y}\right)_{n,n'}\left(\frac{1+\kappa
 \tau ^y}{2}e^{-i t 2\lambda  V_{q,p}\tau ^y}\right)_{p,p'}\left(\frac{1+\lambda  \tau ^y}{2}\right)_{q,q'}\\
 \times &\langle  e^{\frac{1}{4}\gamma
 \left(\mu  A_m+\nu  A_n+\kappa  A_p+\lambda  A_q\right)\gamma }\gamma _{m'} \gamma _{n'} \gamma _{p'} \gamma _{q'}\rangle.
 \end{aligned}
\end{equation}
By denoting $\mu  A_m+\nu  A_n+\kappa  A_p+\lambda  A_q\equiv A$, we can factorize the term similarly and evaluate the effective correlation function part by Wick's theorem:
\begin{equation}
\langle \psi _{\{u\}}|e^{\frac{1}{4}\gamma ^TA \gamma } \gamma _m \gamma _n\gamma _p \gamma _q|\psi _{\{u\}}\rangle =\langle \text{vac}|e^{\frac{1}{4}\zeta
^TQ A Q^T \zeta } \gamma _m \gamma _n\gamma _p \gamma _q|\text{vac}\rangle =-Z\left(\tilde{\Gamma }_{m,n}\tilde{\Gamma }_{p,q}-\tilde{\Gamma }_{m,p}\tilde{\Gamma
}_{n,q}+\tilde{\Gamma }_{m,q}\tilde{\Gamma }_{n,p}\right),
\end{equation}
where
\begin{equation}
Z\equiv \langle \text{vac}|e^{\frac{1}{4}\zeta ^TQ A Q^T \zeta } |\text{vac}\rangle ,
\quad
\tilde{\Gamma }_{m,n}\equiv \frac{\langle \text{vac}|e^{\frac{1}{4}\zeta ^TQ A Q^T \zeta } i\gamma _m \gamma _n|\text{vac}\rangle }{\langle \text{vac}|e^{\frac{1}{4}\zeta
^TQ A Q^T \zeta }|\text{vac}\rangle }
= \sum_{i,j}
Q_{i,m}\frac{\langle \text{vac}|e^{\frac{1}{4}\zeta ^TQ A Q^T \zeta } i\zeta _i \zeta _j|\text{vac}\rangle }{\langle
\text{vac}|e^{\frac{1}{4}\zeta ^TQ A Q^T \zeta }|\text{vac}\rangle }Q_{j,n}.
\end{equation}
Likewise, the Loschmidt amplitude $Z$ and the effective two-point correlation function $\tilde{\Gamma}$ can be evaluated using the general result we are going to derive in the next section. 

In numerical computation, although each four-Majorana-fermion correlation function with fixed $m,n,p,q$ reduces to effective Gaussian evolution and can be efficiently calculated, there are $\sim O(N^4)$ number of independent Gaussian evolving trajectories in total, which is a huge computation complexity to keep track of. Fortunately, it is easy for parallelization. Our computation amounts to $N=16\times 4$ unit-cells, which accounts for 128 spins, being averaged over 200 randomly generated disorder samples. For completeness, here we show the slices of correlation growth in fixed distances, and compare the interacting case with non-interacting case, for both isotropic coupling and anisotropic coupling respectively, see Fig.~\ref{fig:J1interLightCone}. 

\begin{figure}[h] 
   \centering
   \includegraphics[width=.6\columnwidth]{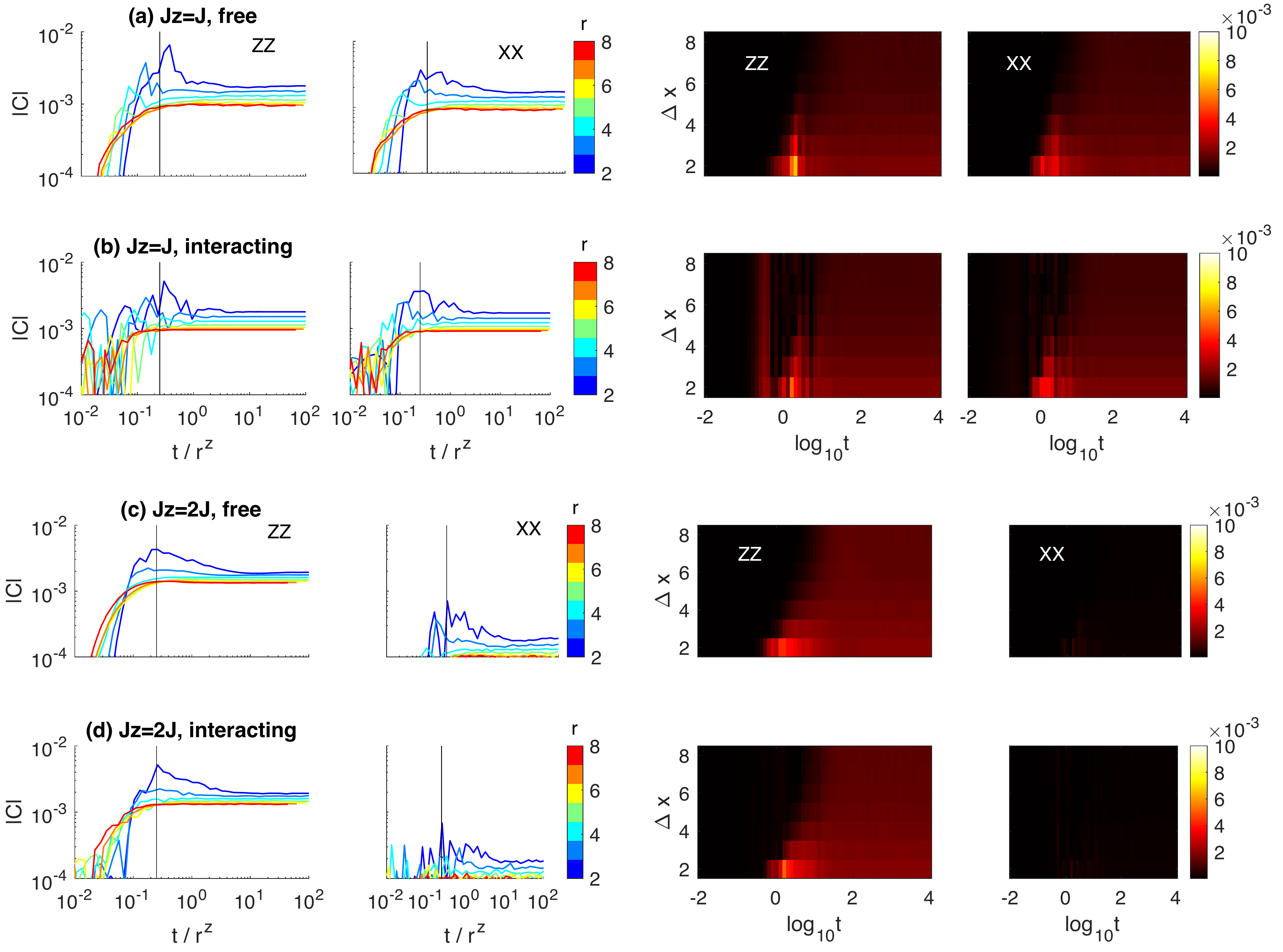} 
   \caption{Correlation growth at fixed distances and the spatiotemporal profiles. (a)(b) isotropic coupling $J_z=J_x=J_y=J, \tilde{h}=0.25J$, using $z=2.4$ to rescale the time axis; (c)(d) anisotropic coupling $J_z=2J, J_x=J_y=0.5J, \tilde{h}=0.25J$. $L_x=16, L_y=4$, using $z'=2.6$ to rescale the time axis. 10000 disorder samples for free fermion case while 200 disorder samples for the interacting scenario.}
   \label{fig:J1interLightCone}
\end{figure}

\subsection{Extended Majorana correlation function}

In this section we are going to derive an independent formula for a generic complex extended Majorana correlation function as below:
\begin{equation}
\begin{aligned}
&\langle \text{vac}|e^{\frac{ 1}{4}\gamma  A \gamma }i\gamma \gamma^T |\text{vac}\rangle =\langle \text{vac}|e^{\frac{ 1}{4}\gamma  A \gamma
}|\text{vac}\rangle \times \frac{\langle \text{vac}|e^{\frac{ 1}{4}\gamma  A \gamma }i\gamma\gamma^T |\text{vac}\rangle }{\langle \text{vac}|e^{\frac{
1}{4}\gamma  A \gamma }|\text{vac}\rangle }
\equiv Z(A)\times \Gamma (A),\\
&Z(A)=(-1)^{\frac{N(N-1)}{2}}\det (Q)\text{Pf}\left(Q^Ti\tau^yQ-i\tau^y\tanh \left(\frac{a_n}{2}\right)\right)\left(\prod
_n\cosh \left(\frac{a_n}{2}\right)\right),
\quad
\Gamma(A)= i\left(1+\tau ^y+\left(\tau ^z-i \tau ^x\right)\otimes K\right),
\end{aligned}
\end{equation}
where the $|\text{vac}\rangle$ is defined by $\langle \text{vac} | i\gamma\gamma^T | \text{vac}\rangle = i(1+\tau^y)$, $\gamma$ stands for a column vector of Majorana fermion operators. The complex correlation function is factorized into a product of a Loschmidt amplitude and an effective correlation function. In general, the input argument $A$ can be any generic complex antisymmetric matrix that has the canonical form, and $K$ is the matrix that would appear as the Gaussian exponent coupled to the pairing term in the Balian-Brezin decomposition:
\begin{equation}
Q^TA Q=\tau ^y\otimes (\oplus_n a_n),\qquad
K=\left(\left(
\begin{array}{cc}
 -i & 1 \\
\end{array}
\right)e^A\left(
\begin{array}{c}
 i \\
 1 \\
\end{array}
\right)\right)^{-1}\left(\left(
\begin{array}{cc}
 -i & 1 \\
\end{array}
\right)e^A\left(
\begin{array}{c}
 -i \\
 1 \\
\end{array}
\right)\right).
\end{equation}
This formula can be applied to evaluate multiple expressions in the former section, where we turned back to the original fermion basis $\zeta$, and regauge the initial state to be a fermion vacuum of $\zeta$. The Pfaffian can be numerically evaluated efficiently using the algorithms and codes provided by Ref.~\cite{Wimmer2011}. 

Detailed derivation for this formula are given as follows. To evaluate the Loschmidt amplitude without sign ambiguity, we first derive a generic formula for the expectation of generic Majorana fermion Gaussian
operator over the fermion vacuum, by virtue of the Baker-Campbell-Hausdorff formula for tracing out sequence of Majorana Gaussian operators as relevant in the generalized Levito's formula $\text{Tr} (e^{\frac{1}{4}\gamma  A \gamma} e^{\frac{1}{4}\gamma  B \gamma} )=\sqrt{\det{(1+e^A e^B)}}$~\cite{Klich2002,Klich2014}:
\begin{equation}
\begin{aligned}
Z(A)&\equiv \langle \text{vac}|e^{\frac{ 1}{4}\gamma  A \gamma }|\text{vac}\rangle 
=\text{lim}_{\beta\to\infty}\text{Tr}\left(e^{\frac{-\beta }{4}\gamma  \tau ^y\gamma }e^{\frac{
1}{4}\gamma  A \gamma }\right)/\text{Tr}\left(e^{\frac{-\beta }{4}\gamma  \tau ^y\gamma }\right)
=\text{lim}_{\beta\to\infty}\sqrt{\det \left(\frac{1}{1+e^{-\beta  \tau
^y}}+\frac{1}{1+e^{\beta  \tau ^y}}e^A\right)}\\
&=\sqrt{\det \left(\frac{1+\tau ^y}{2}+\frac{1-\tau ^y}{2}e^A\right)}
=\sqrt{\det \left(\frac{1+\tau
^y}{2}e^{-A/2}+\frac{1-\tau ^y}{2}e^{A/2}\right)}
=\sqrt{\det \left(\cosh \left(\frac{A}{2}\right)-\tau ^y\sinh \left(\frac{A}{2}\right)\right)}\\
&=\sqrt{\det
\left(1-\tau ^y\tanh \left(\frac{A}{2}\right)\right)}\sqrt{\det \left(\cosh \left(\frac{A}{2}\right)\right)}
=\sqrt{\det \left(
\begin{array}{cc}
 i\tau^y & -1 \\
 1 & i\tanh\left(\frac{A}{2}\right) \\
\end{array}
\right)}\left(\prod _n\cosh \left(\frac{a_n}{2}\right)\right)\\
&= \text{Pf}\left(
\begin{array}{cc}
 i\tau^y & -1 \\
 1 & i\tanh\left(\frac{A}{2}\right) \\
\end{array}
\right)\left(\prod _n\cosh \left(\frac{a_n}{2}\right)\right)
=\text{Pf}\left(i\tau^y\right)\text{Pf}\left(i\tau^y-i\tanh\left(\frac{A}{2}\right)\right)\left(\prod
_n\cosh \left(\frac{a_n}{2}\right)\right)\\
&=(-1)^{\frac{N(N-1)}{2}}\det (Q)\text{Pf}\left(Q^Ti\tau^yQ-i\tau^y\tanh \left(\frac{a_n}{2}\right)\right)\left(\prod
_n\cosh \left(\frac{a_n}{2}\right)\right).
\end{aligned}
\end{equation}
Notice that when getting rid of the square root there could be a sign factor, which is fixed to be positive in this case by analytic continuity such that $Z(A)\to 1$ when $A\to 0$, as long as $Z(A)$ remains analytic along the path of A variation and does not trespass a zone with condensed Lee-Yang-Fisher zeroes in thermodynamic limit. In finite size system, $Z(A)$ is a finite order polynomial complex function where isolated Fisher zeroes can always be avoided by infinitesimal variation. Notice that the final expression
is rather complex compared to a simple $\sqrt{\det}$ at the beginning lines of derivation, but being rewritten in terms of Pfaffian has the advantage of being free from sign ambiguity~\cite{Robledo2009}. Otherwise, summation over different trajectories with a sign ambiguity would lead to a sign problem like in quantum Monte Carlo calculations typically for fermion systems.

Next we derive the generic correlation matrix defined with respect to an effective Gaussian operator $|\text{vac}\rangle\langle \text{vac}|e^{\frac{ 1}{4}\gamma  A \gamma }\equiv \lim_{\beta\to\infty}e^{-\beta\frac{ 1}{4}\gamma  \tau^y \gamma }e^{\frac{ 1}{4}\gamma  A \gamma }$:
\begin{equation}
-i\Gamma(A)=\frac{\langle \text{vac}|e^{\frac{ 1}{4}\gamma  A \gamma }\gamma \gamma^T |\text{vac}\rangle }{\langle \text{vac}|e^{\frac{ 1}{4}\gamma
 A \gamma }|\text{vac}\rangle }
 =\lim_{\beta\to\infty} 2\left(1+e^{-\beta  \tau ^y}e^A\right)^{-1}=2\left(\frac{1+\tau ^y}{2}+\frac{1-\tau ^y}{2}e^A\right)^{-1}\frac{1+\tau^y}{2},
\end{equation}
by applying the general formula for correlation function $\text{Tr}(e^{\frac{1}{4}\gamma B \gamma} \gamma \gamma^T) /\text{Tr}(e^{\frac{1}{4}\gamma B \gamma}) = 2(1+e^B)^{-1}$ which can be easily proved when B can be decomposed by orthogonal transformation to a canonical form for disentangled Majorana fermion pairs. 
It involves a projector that can be further simplified in the Bogoliubov-de-Gennes(BdG) basis:
\begin{equation}
-i\Gamma_{\text{BdG}}=-i\frac{1}{\sqrt{2}}\left(
\begin{array}{cc}
 i & 1 \\
 -i & 1 \\
\end{array}
\right)\Gamma \frac{1}{\sqrt{2}}\left(
\begin{array}{cc}
 -i & i \\
 1 & 1 \\
\end{array}
\right)=2\left(\frac{1+\tau ^z}{2}+\frac{1-\tau ^z}{2}e^{A_{\text{BdG}}}\right)^{-1}\frac{1+\tau ^z}{2}=2\left(
\begin{array}{cc}
 1 & 0 \\
 -T_{22}^{-1}T_{21} & 0 \\
\end{array}
\right),
\end{equation}
where we denote $T\equiv e^{A_{\text{BdG}}}$ and the subscript labels particle-hole blocks. Notice that the above can also be directly derived
by using the Balian-Brezin decomposition for $\langle \text{vac}|e^{\frac{ 1}{4}\gamma  A \gamma }c^{\dagger } c^{\dagger }|\text{vac}\rangle =\langle
\text{vac}|e^{c K c}c^{\dagger } c^{\dagger }|\text{vac}\rangle$, where the anti-symmetric matrix
\begin{equation}
K\equiv T_{22}^{-1}T_{21}=\left(\left(
\begin{array}{cc}
 -i & 1 \\
\end{array}
\right)e^A\left(
\begin{array}{c}
 i \\
 1 \\
\end{array}
\right)\right)^{-1}\left(\left(
\begin{array}{cc}
 -i & 1 \\
\end{array}
\right)e^A\left(
\begin{array}{c}
 -i \\
 1 \\
\end{array}
\right)\right).
\end{equation}
Turning back to the Majorana fermion basis, we have
\begin{equation}
-i\Gamma=1+\tau ^y+\left(\tau ^z-i \tau ^x\right)\otimes K,
\end{equation}
that completes the derivation. 

Finally we also comment that the expression above can also be consistently obtained by using the general formula of tracing out the product of two generic Gaussian operators, and the formula of the product of two Gaussian correlation functions~\cite{Fagotti_2010} up to some further simplification. Namely, for any generic square skew-symmetric matrices $A$ and $B$, 
\begin{equation}
\begin{split}
&\frac{\text{Tr}(e^{\frac{1}{4}\gamma^T A\gamma}e^{\frac{1}{4}\gamma^T B\gamma})}
{\text{Tr}(e^{\frac{1}{4}\gamma^T A\gamma})\text{Tr}(e^{\frac{1}{4}\gamma^T B\gamma})}
=
\frac{1}{2^{\text{dim}(A)}}\text{Pf}(\Gamma_B)\text{Pf}(\Gamma_A+\Gamma_B^{-1}),\\
&\frac{\text{Tr}(e^{\frac{1}{4}\gamma^T A\gamma}e^{\frac{1}{4}\gamma^T B\gamma} 
\frac{i}{2}[\gamma, \gamma^T])}
{\text{Tr}(e^{\frac{1}{4}\gamma^T A\gamma}e^{\frac{1}{4}\gamma^T B\gamma})}
=
1-(1-\Gamma_B)\frac{1}{1+\Gamma_A\Gamma_B}(1-\Gamma_A),\\
&\Gamma_{A} = -\tanh (A/2).
\end{split}
\end{equation} 
Notice that in the general formula here $\Gamma_{A}$ is the purely skew-symmetric correlation matrix with the constant diagonal part being subtracted, slightly different from the ones we derive before. 

\section{Localization length in two dimensions}

We use the standard free fermion method to calculate the localization length in two dimensions by scaling the quasi-1D localization length~\cite{MacKinnon, MacKinnon1983, Markos2006} . The first step is to put the system into a semi-infinite long stripe
with fixed width $L_y$, and calculate the localization length for different energy $\lambda \left(E,L_y\right)$ using the transfer matrix or
recursive Green's function connecting long distance. In a generic quasi 1D system, the retarded correlation connecting site 1 to far enough site
$x$ is generally expected to decay exponentially except on critical point
$\left\|G_{1,x}\right\|\propto e^{\frac{- L_x}{\lambda }}$,
from which one can determine the localization length $\lambda $. To reduce the rounding error, one can recursively generate the correlation function,
rescaling it while recording the norm decaying rate in each step. Averaging the recorded decaying rates gives the smallest Lyapunov exponent $z$ which
is inverse proportional to the localization length $\lambda $, and since the disorder is generated in each step independently, the statistical variance of the Lyapunov exponent is accumulated:
\begin{equation}
\bar{z}=-\frac{1}{L_x}\sum _{x=1}^{L_x}\ln  \frac{\left\|G_{1,x+1}\right\|}{\left\|G_{1,x}\right\|} \equiv \frac{1}{\lambda },
\qquad
\sqrt{\text{var} z} =\frac{1}{L_x} \sqrt{\sum _{x=1}^{L_x}\left(-\ln  \frac{\left\|G_{1,x+1}\right\|}{\left\|G_{1,x}\right\|}\right)^2}\qquad 
\propto \qquad\frac{\bar{z}}{\sqrt{L_x}},
\end{equation}
according to the central limit theorem, given that a finite localization length exists. We'll calculate $L_y=8,16,32,64,128$, and set the convergence threshold
$\sqrt{\text{var} z}/\bar{z} \leq  0.01$ to stop the iteration. For the case that does not meet such criterion up to $L_x=10^6$, we'll
stop the iteration and let loose the threshold and keep the data with $\sqrt{\text{var} z}/\bar{z} \leq  0.1$ as tolerable. In detail, our
free Majorana fermion Hamiltonian matrix is sliced into columns with intra-column and inter-column Hamiltonian coupling (using $r$ to label the unit-cell
and $x$ to label the column,  $n_2$ being the primitive vector along column and $n_1$ across columns):
\begin{equation}
H=\sum _x \left( H_x+V_{x,x+1}+V_{x,x+1}^{\dagger }\right),
\end{equation}
\begin{equation}
H_x=\sum _{r_2}\left(
\begin{array}{cc}
 0 & -i J_z \\
 0 & 0 \\
\end{array}
\right)u_r|r\rangle \langle r|+\left(
\begin{array}{cc}
 0 & -i J_x \\
 0 & 0 \\
\end{array}
\right)|r\rangle \langle r+n_2|+\left(
\begin{array}{cc}
 -i \tilde{h} \tau _r & 0 \\
 0 & 0 \\
\end{array}
\right)|r\rangle \langle r-n_2|+\left(
\begin{array}{cc}
 0 & 0 \\
 0 & -i \tilde{h} \tau _r \\
\end{array}
\right)|r\rangle \langle r+n_2|+h.c.,
\end{equation}
\begin{equation}
V_{x,x+1}=\sum _{r_2} \left(
\begin{array}{cc}
 0 & 0 \\
 i J_y & 0 \\
\end{array}
\right)|r\rangle \langle r-n_2|+\left(
\begin{array}{cc}
 -i \tilde{h} & 0 \\
 0 & i \tilde{h} \\
\end{array}
\right)|r\rangle \langle r|\\
\\
+ \left(
\begin{array}{cc}
 i \tilde{h} & 0 \\
 0 & 0 \\
\end{array}
\right) u_r|r\rangle \langle r-n_2|+ \left(
\begin{array}{cc}
 0 & 0 \\
 0 & -i \tilde{h}  \\
\end{array}
\right)|r\rangle \langle r-n_2|u_{r+n_1-n_2}.
\end{equation}
The correlation matrix connecting site 1 and site $x$ follows from the recursive relation derived from Dyson equation
\begin{equation}
G_{0,x}=G_{0,x-1} V_{x-1,x}G_{x,x},
\qquad
G_{x,x}^{-1} = E-H_x-V_{x-1,x}^{\dagger }G_{x-1,x-1}V_{x-1,x},
\end{equation}
by initiating the left semi-infinite chain $G_{0,1}=1, G_{1,1}=E-H_x$.

The second step is to fit the data into a one parameter scaling ansatz: 
\begin{equation}
\frac{\lambda \left(E,L_y\right)}{L_y}=f\left(\frac{\xi (E)}{L_y}\right),
\end{equation}
where $\xi (E)$ is the localization length we want. Numerically, this can be achieved by minimizing the variance of $\ln\xi(E) - \ln L_y$ for the interpolated data points $\left(E,L_y\right)$ that shares identical $\lambda \left(E,L_y\right)/L_y$~\cite{MacKinnon1983}. The scaling function obtained by data collapse
is shown as in Fig.~\ref{fig:locLengthScalCollapse}.

\begin{figure}[h] 
   \centering
   \includegraphics[width=9cm]{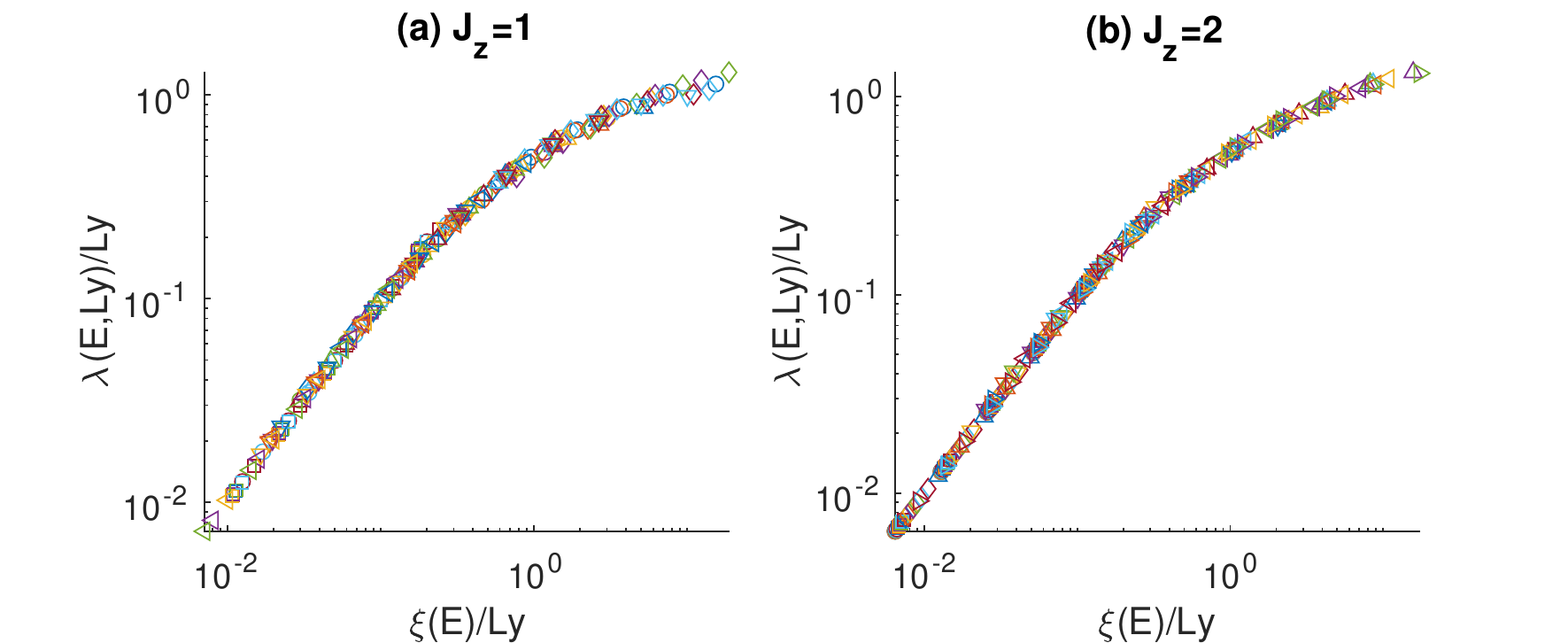} 
   \caption{Data collapse for localization length scaling function. In large width limit the scaling function is expected to approach a linear regime. (a) $J_z=J_x=J_y=J, \tilde{h}=0.25J, L_y=8,16,32,64,128, L_x\leq 10^6$. (b) $J_z=2J, J_x=J_y=0.5J, \tilde{h}=0.25J, L_y=8,16,32,64,128, L_x\leq 10^6$. Data with relative deviation of smallest Lyapunov exponent greater than 0.1 is excluded from the plot. }
   \label{fig:locLengthScalCollapse}
\end{figure}

\section{Chern number in disorder}

In this section we work in a typical disordered gauge configuration in the absence of fermion interaction, and resolve the localized/delocalized nature of the single-particle eigenstate wave-functions by calculating the Chern number. In a disordered configuration without translation
and conservation of momentum, the Hamiltonian is no longer block diagonal in momentum space, so that the protypical form of the TKNN Chern number formula~\cite{Thouless} with
summation over Brillouin zone can no longer be simply used. But there are in general two ways to remedy this problem. 

One way is to generalize the
Brillouin zone using the concept of non-commutative geometry, which manifest in replacing summation by matrix trace in calculation, and is guaranteed
to be convergent and topological if in the presence of a mobility gap~\cite{Bellissard1994,Kitaev2005,Bianco2011}. The formula we need to calculate is 
\begin{equation}
C=\frac{2\pi }{i}\frac{1}{N}\text{Tr}(P[[i x,P], [i y,P]])=2\pi  i\frac{1}{N}\text{Tr}([P x P, P y P]),
\end{equation}
where $P$ is the spectral projector to be concretely defined below, the real space continuum coordinate operator $x(y)$ is to be expanded by the lattice coordinate in a properly defined series~\cite{Prodan2010} such that the finite size error is exponentially small. Our numerical result in the main text is based on this method. 

The other way is to twist the boundary phase instead~\cite{Niu}, which takes the spirit
of Laughlin's gedanken experiment of flux pumping, and can be understood as generating a super-lattice with effective quasi-momentum. In numerical calculation for a finite size lattice of discretized twisting boundary phase, it has great advantage to regularize the problem into the form of a lattice gauge theory, whose global flux over the twisting phase space is guaranteed to be quantized~\cite{Fukui2005}:
\begin{equation}
C=\frac{1}{2\pi }\sum _{\phi}\arg \left(U_{\phi }^xU_{\phi +d\phi_x}^yU_{\phi +d\phi_y}^{x*}U_{\phi }^{y*}\right),\qquad 
U_{\phi }^{x(y)}=\left\langle
\psi _{\phi }|\psi _{\phi +d\phi_{x(y)}}\right\rangle ,
\end{equation}
where $\phi$ is the twisting boundary phase, $|\psi_\phi\rangle$ is the corresponding single particle or manybody wave-function. The other advantage of this approach is that for a free fermion problem it results in eigen-state resolved Chern numbers, from which one can determine the density of extended states~\cite{Arovas,Halperin1982,Huo}, as shown in Fig.~\ref{smfig:ChernNumberExtDOS}. 

\begin{figure}[h] 
   \centering
   \includegraphics[width=8cm]{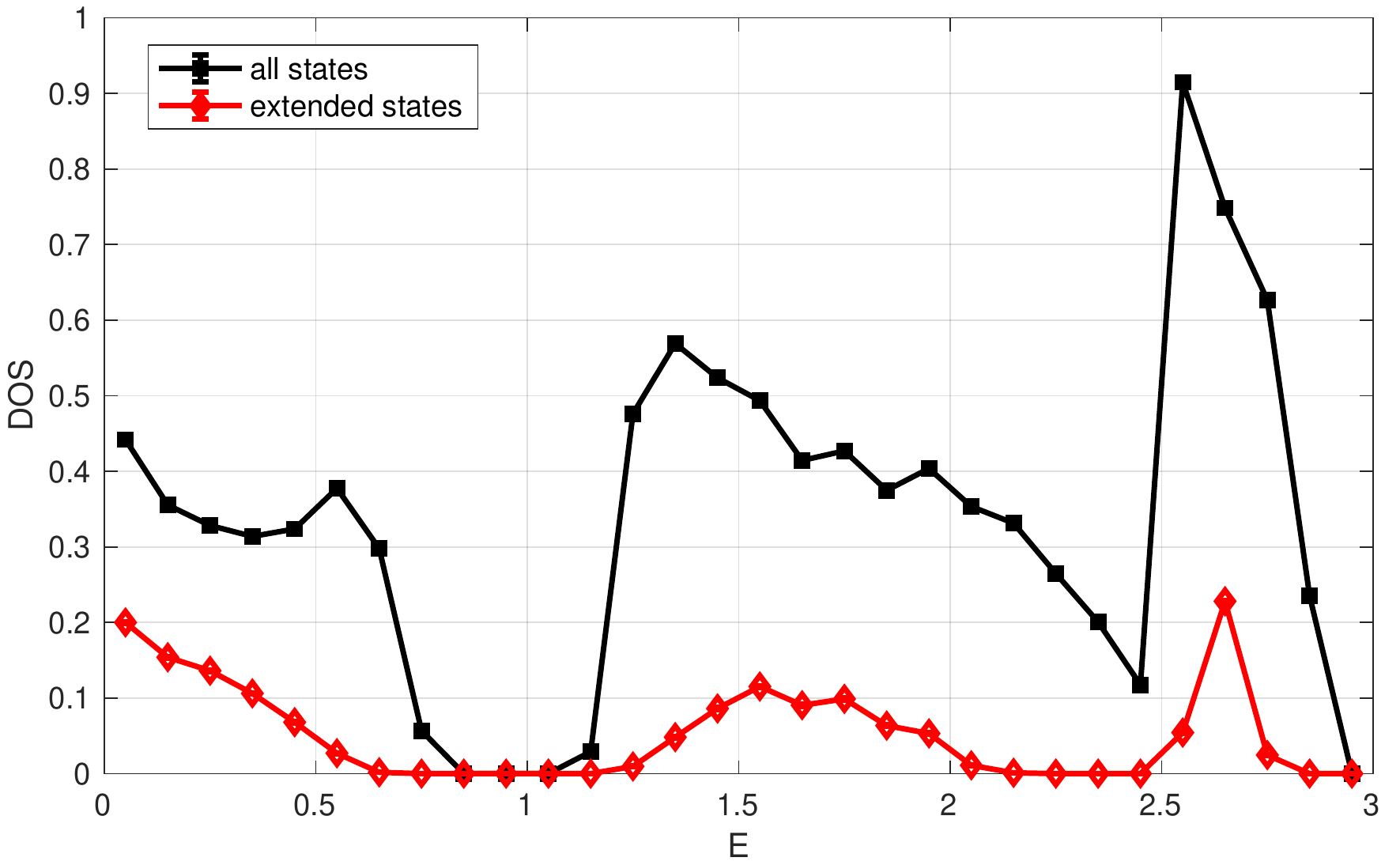} 
   \caption{Black lines: density of states from exact diagonalization in the absence of twisted phase. Red lines: density of extended single-particle states that carry nonzero Chern number, calculated in twisting boundary phase space. Parameters: $J_z=J_x=J_y=J, \tilde{h}=0.25J, L_x=L_y=40$, 1000 disorder samples. Twisted phase space is discretized into $20\times20$ square lattice. }
   \label{smfig:ChernNumberExtDOS}
\end{figure}

In the following we briefly review the technical details for these two methods. To avoid confusion, all the notations work only within the following  sub-section independently and does not apply to the rest of the supplemental material. 

\subsection{Calculation using real space formula}

Even without translation symmetry, one can still perform a Fourier transform and define momentum, although it is no longer good quantum number, because
the single particle Hamiltonian in momentum space carries off-diagonal elements connecting different momenta. Naively, one can simply generalize
the summation over momentum to matrix trace over of the spectral projector. Concretely, first let's denote the single particle spectral projector to the ground state in real space basis:
$P=\frac{1-\text{sgn}(H)}{2}$,
where sgn($H$) is obtained by flattening the spectrum of the single particle Hamiltonian $H$ but the sign. The spectral projector in momentum space and its derivative are
\begin{equation}
\tilde{P}=\frac{1}{N}e^{i k\cdot r}P e^{-i k\cdot r},
\qquad 
\partial _k\tilde{P}=\frac{1}{N}e^{i k\cdot r}[i r,P] e^{-i k\cdot r}.
\end{equation}
One may generalize the Chern number formula
\begin{equation}
\begin{aligned}
C=\frac{1}{2\pi i}\int dk_xdk_y\text{Tr}\left(P(k)\left[\partial _{k_x}P(k), \partial _{k_y}P(k)\right]\right)
&\rightarrow \frac{2\pi }{i}\frac{1}{N}\text{Tr}\left(\tilde{P}\left[\partial _{k_x}\tilde{P}, \partial _{k_y}\tilde{P}\right]\right)
=\frac{2\pi }{i}\frac{1}{N}\text{Tr}\left(\frac{1}{N}e^{i k\cdot r}P[[i x,P], [i y,P]]e^{-i k\cdot r}\right)\\
&=\frac{2\pi }{i}\frac{1}{N}\text{Tr}(P[[i x,P], [i y,P]])
=2\pi  i\frac{1}{N}\text{Tr}([P x P, P y P]).
\end{aligned}
\end{equation}
In the end, it is simply a commutator of the real space coordinate operator being projected into the Fermi sea. One could immediately verify that
when $P=1\Rightarrow  C\propto[x,y]=0$, the contribution over the total Hilbert space is guaranteed to vanish. 
The essential questions are whether it converges to integer and whether it is of topological nature. First, it was elaborated early by Bellisard using the concept of non-commutative geometry~\cite{Bellissard1994}, and later Kitaev also gave an intuitive argument for the topological nature of the 2-current of the spectral projector, and its quantization as the flow of a quasi-diagonal unitary matrix reminiscent of the Laughlin's flux pumping gedanken experiment $F\left(e^{i 2\pi  P x P}\right)=\left.\text{Tr}\left[e^{i 2\pi  P x P}P y P e^{-i 2\pi  P x P}-P y P\right]\right/N=\int
_0^{2\pi }\text{Tr}\left[\partial _{\phi }e^{i \phi  [P x P, ]}P y P\right]d\phi /N=2\pi  i [P x P, P y P]/N$~\cite{Kitaev2005}. Note that the quasi-diagonal condition
of the unitary matrix is readily satisfied when there is a spectral or mobility gap so that correlation exponentially decays. Note also that Bianco et.al. also derived a similar formula from the linear response theory for continuum real space system~\cite{Bianco2011}, and they even took out the local contribution in each unit-cell i.e. the diagonal entry of the commutator $2\pi i[PxP,PyP]$ and defined a so-called local Chern marker, which in a clean system is equivalent to the total Chern number. However, one should notice that the trade
of $k$-derivative with commutator of coordinate $\partial _k\to  [i r,]$ is exact only at thermodynamic limit $N\to \infty $ such that $\Delta k\to 0$ can be infinitesimally
small. Prodan et.al. proposed an efficient numerical algorithm to exponentially suppress the finite size error in this line as follows~\cite{Prodan2010}. In a finite size system with $L$ units along $x$-direction, one could approximate the real-space coordinate by an expansion
$x\simeq \sum _{n=1}^{L/2}a_n\left(e^{i n \frac{2\pi }{L} x}-c.c.\right)$.
As what we need is not to cover the whole lattice coordinate, but to approach the continuum limit of small $x$, a direct discrete Fourier
transform for the lattice coordinate is apparently not the optimal choice, which yields polynomial $O(1/L^2)$ error. One optimal solution
is chosen by solving the linear matrix equation $\sum _{n=1}^{L/2}a_n n ^{2j-1}=\frac{L}{2\pi }\delta _{j,1}, (j=1,\cdots,L/2)$
such that the deviation between the continuous $x$ and the expanded series is exactly 0 up to $L/2$-th order, leading to exponentially small error with the system size: $O\left(\frac{1}{L}\right)^L$. As coordinate operator generates the translation of momentum, we have
\begin{equation}
[i x,P]=e^{-i k\cdot r}\left(\partial _k\tilde{P}\right)e^{i k\cdot r}=\sum _{n=1}^{L/2}a_n\left(e^{i \frac{2\pi }{L}n\cdot r}P e^{-i \frac{2\pi
}{L}n\cdot r}-e^{-i \frac{2\pi }{L}n\cdot r}P e^{i \frac{2\pi }{L}n\cdot r}\right)+O\left(\frac{1}{L}\right)^L.
\end{equation}

\subsection{Calculation by twisting boundary phase}

It is akin to the momentum space formula for the Chern number~\cite{Thouless,Halperin1982,Niu}. By discretizing the twisted boundary phase space into $M_x\times M_y$ sites, it
is equivalent to treat the $L_x\times L_y$ finite-size system as a giant unit-cell and generate a periodic super-lattice of size $L_x\times L_y\times
M_x\times M_y$, thus effectively generating a mini-Brillouin-zone(BZ) for the quasi-momentum in unit of $2\pi \left/M_{x(y)}\right.$. In the mini-BZ, the $L_x\times
L_y$ large unit cell folds into $L_x\times L_y$ mini-bands. Since the twisting phase $\phi $ , conjugate to the real space coordinate, plays the
role of a quasi-momentum ranging between $[0,2\pi /L]$ ~\cite{Niu}:
\begin{equation}
C=\frac{1}{2\pi i}\int _{\partial \text{BZ}}d\overset{\to }{\phi }\cdot \langle \psi |\partial \psi \rangle =\frac{1}{2\pi }\int _{\text{BZ}}d\phi _xd\phi
_yF_{\phi },
\qquad
F_{\phi }=i\left\langle \partial _{\phi _x}\psi (\phi )|\partial _{\phi _y}\psi (\phi )\right\rangle +h.c.
\end{equation}
where $\psi $($\phi $) can either be the single particle eigenstate or manybody eigenstate for the Hamiltonian with twising boundary phase $\phi$. Physically, the twisting boundary phase also concurs with Laughlin's
flux pumping gedanken experiment. In a finite-size lattice system, one could regularize the above formula like regularizing the continuum gauge theory
into a lattice gauge theory~\cite{Fukui2005}. Namely, regularize the local flux integrated over a plaquette as the Wilson loop of the $U(1)$ link
variable, which is physically the parallel transportation of the wave-function:
\begin{equation}
C=\frac{1}{2\pi }\sum _{\phi}\arg \left(U_{\phi }^xU_{\phi +d\phi_x}^yU_{\phi +d\phi_y}^{x*}U_{\phi }^{y*}\right),\qquad 
U_{\phi }^{x(y)}=\left\langle
\psi _{\phi }|\psi _{\phi +d\phi_{x(y)}}\right\rangle .
\end{equation}
where arg is defined as taking the phase angle from the principal branch $(-\pi , \pi ]$. Notice that the parallel transporter circulating the BZ should be equivalent to identity due to the periodicity (close manifold without boundary), which means the total flux must be quantized in units of $2\pi$, therefore guaranteeing the quantization of Chern number defined on this lattice. Whether this global flux coincides with the continuum limit is a question. While there is no limitation to the flux on each plaquette in the continuum limit as a noncompact gauge theory, there is $2\pi$ ambiguity of flux on each plaquette in the lattice gauge theory with compact gauge group. This is the source of the discrepancy between the lattice and the continuum limit. Generally, when the phase grid of the lattice is fine enough such that the corresponding flux on each plaquette in continuum limit is controlled within the principal branch window, the finite size result should unambiguously concurs with the continuum limit. Assuming a smooth Berry curvature configuration,
by uniformly distributing the Berry flux onto each plaquette approximately, one can estimate a lower-bound for the required size of the phase space: $M_xM_y>2|C|$.
As we know, for a low-energy Dirac fermion with light mass, the Berry curvature is sharply peaked around the Dirac point, but the distribution is flattened when the mass is large. 

The quantization of Chern number in continuum limit relies on the adiabatic evolution of the ground state wave-function circulating around the twisting phase space.
In other words, it also requires a mobility gap at the Fermi-level, or a (exponentially) fast decay of the correlation function, which excludes
the situation with Fermi-level crossing a band in a clean system. In cases with band overlapping or level crossing, the Berry curvature could have
singularity. However, in a generally disordered finite size system, this generally holds because either the states near the Fermi-level is localized
without contribution to Berry curvature, or the delocalized states near Fermi-level are generally experiencing level repulsion due to scattering.
Notice that the Bloch state in clean system without momentum scattering and manybody interaction is a very special case in this sense. 

In practice, one can either use the manybody ground state (such as the slater determinant for free fermion system) to construct the link variable, or resolve the Berry phase for each single particle eigen-state, which gives biproduct of determining the localization nature of a wave-function~\cite{Arovas,Halperin1982}. Further, one can perform scaling analysis for the density of extended states versus
the total states to deduce the existence of extended states in thermodynamic limit that escapes localization~\cite{Huo}.

\section{Zero energy thermal metallic state}

In this section we show numerical verification for the delocalized nature of the single particle mode at zero energy for the isotropic coupling model for larger system size, which is qualitatively consistent with the infinite temperature limit of a moderate size numerical calculations in Ref.~\cite{Self2018}. To resolve the density of states at extremely small energy, we use the standard kernel polynomial method~\cite{Weisse2005}, to expand the density of states, defined as a collection of $\delta $ functions, in the orthogonal basis of Chebyshev polynomials, 
\begin{equation}
\rho (E) = \frac{1}{\pi \sqrt{1-\left(E\left/E_{\max }\right.\right){}^2}}\left(1+2\sum _{n=1}^{\infty }\text{Tr}\left(T_n\left(H\left/E_{\max }\right.\right)\right)T_n\left(E\left/E_{\max
}\right.\right)\right).
\end{equation}
where $T_n$ is the first kind of Chebyshev polynomials, and the coefficient at each order can be evaluated iteratively in terms of sparse matrix. 10 random complex states are initiated to evaluate the trace of matrix stochastically. In this way we can go to very large system
size with finer energy resolution. When truncated at finite orders, a Jackson kernel is attached to damp the Gibbs oscillations, resulting in a smooth regular function. Results are shown in Fig.~\ref{smfig:DOSMultiFractal}ab. A logarithmic divergent behaviour close to zero energy is visible, consistent with the theoretical prediction~\cite{Senthil1999}. 

To verify the weak multi-fractal nature of the single fermion wave-function at zero energy, we first look at the fractal dimension of the generalized
inverse participation ratio~\cite{Evers2007}:
\begin{equation}
D_q=\frac{1}{1-q}\frac{\ln  \sum _rp(r)^q}{\ln  L},
\end{equation}
where $p(r)\equiv \left|\psi (r)|^2\right.$ is the zero energy single particle wave-function probability distribution, $L$ is the coarse grained linear system size. In practical numerical calculation, to reduce the interference of the lattice length scale, we typically do a coarse graining for $p(r)$, where $r$ is the coarse grained coordinate. To calculate the singularity spectrum $f(\alpha)$ that is related to $D_q$ by the Legendre transform $D_q(1-q)=f(\alpha )-\alpha  q, f'(\alpha )=q$, we first
define a set of normalized weight $\mu _q(r) = p(r)^q/\sum _rp(r)^q $. Then the scaling dimension of $q$-th moment weighted average probability, and the associated volume scaling dimension are respectively given by
\begin{equation}
\alpha (q)=\frac{\sum _r\mu _q(r)\ln  p(r)}{-\ln  L},\qquad
f(\alpha (q))=\frac{\sum _r\mu _q(r)\ln  \mu _q(r)}{-\ln  L}.
\end{equation}
It can be checked that $f(\alpha (q=0))=2$. Similar to a 2+$\epsilon $ dimensional metal near the Anderson localization, our results show weak multi-fractal bahaviour $D_q=2-\gamma  q$, when $\gamma  q \ll 2$, and $\alpha (q=0)=d-\gamma$, as shown in Fig.~\ref{smfig:DOSMultiFractal}cd. 

\begin{figure}[H] 
   \centering
   \includegraphics[width=9cm]{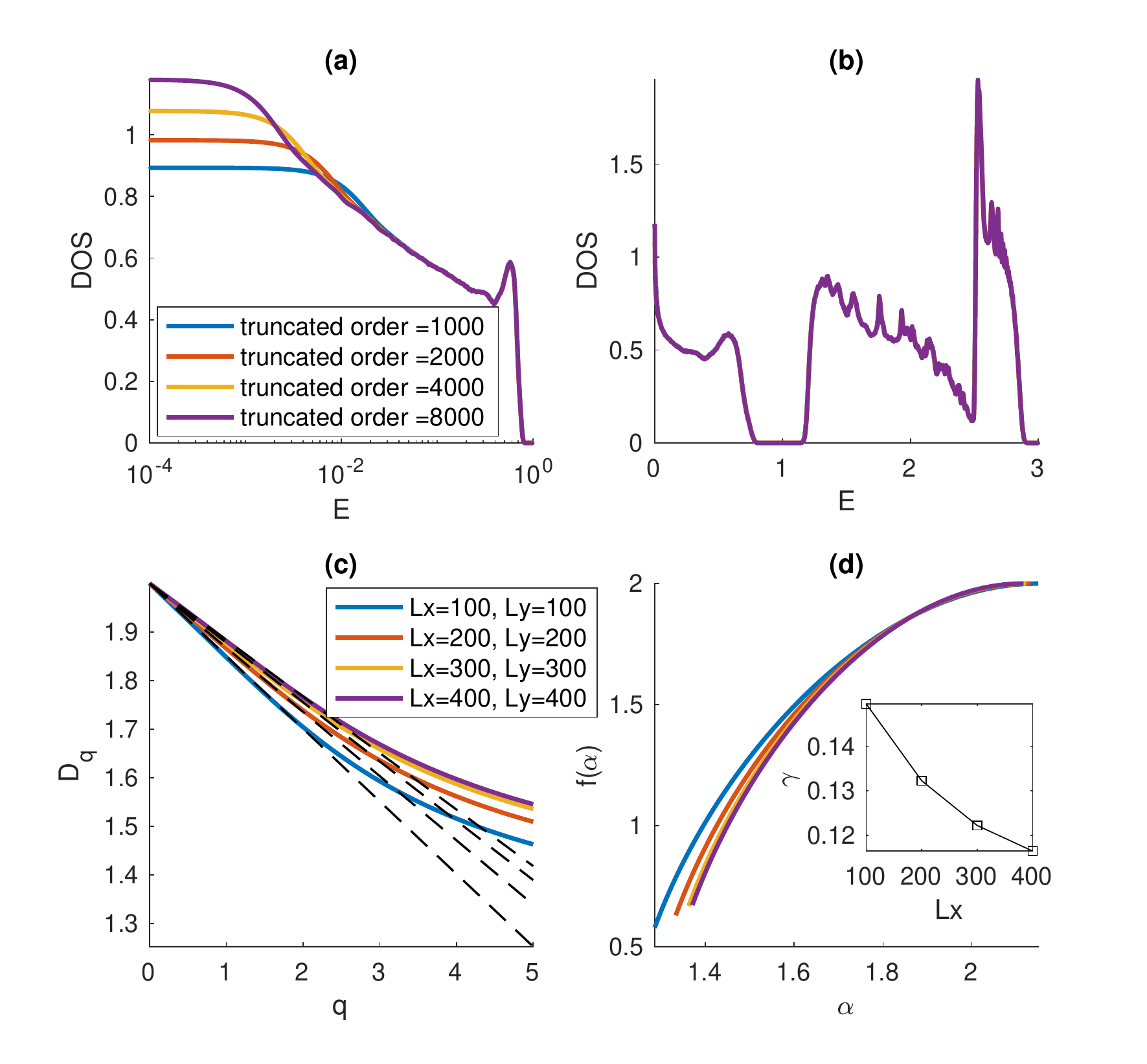} 
   \caption{(a)(b) Density of states (DOS) calculated using kernel polynomial method with different truncated orders. $L_x=L_y=100$. The large scale DOS is consistent with the exact diagonalization result in Fig.~\ref{smfig:ChernNumberExtDOS}. (c) Fractal dimensions of the generalized inverse participation ratio defined out of $q$-th moment of zero energy single particle wave-function probability distribution. Dashed lines sketch the weak multifractal behaviour: $D_q=2-\gamma q$. (d) Singular spectrum of wave-function intensity distribution. Inset shows the finite size dependence of $\gamma$. Parameters: $J_z=J_x=J_y=J, \tilde{h}=0.25J$. }
   \label{smfig:DOSMultiFractal}
\end{figure}
}

\end{document}